\documentclass[leqno, 12pt]{article}
\usepackage{oxford3}
\usepackage{kokoszka}
\usepackage{graphicx}
\usepackage{amsfonts}
\usepackage{amsmath}
\usepackage{amssymb}
\usepackage{enumitem}
\usepackage{color}
\usepackage{graphicx}

\renewcommand{\baselinestretch}{1.02}
\newcommand{\mi}{\mathrm{i}} 
\newcommand\mbR{{\mathbb R}}
\newcommand\mbZ{{\mathbb Z}}

\newcommand\cQ{{\mathcal Q}}

\newcommand{\se}[1]{\textit{\scriptsize #1}}
\newcommand{\gpu}[1]{\textit{#1}}
\allowdisplaybreaks

\usepackage{xcolor}
\usepackage{hyperref}
\hypersetup{
    colorlinks=true,
    linkcolor=black,       
    citecolor=black,        
    filecolor=magenta,    
    urlcolor=cyan         
}

\begin{document}

\title{Deep learning estimation of the spectral density of
functional time series on large domains }

\author{
Neda Mohammadi\\
{\small University of Texas   El Paso} 
\and
Soham Sarkar\\
{\small Indian Statistical Institute}
\and
Piotr Kokoszka\footnote{Correspondence to: Piotr Kokoszka, Colorado State
University, Fort Collins, CO 80523-1877, USA.
\newline  Email: Piotr.Kokoszka@colostate.edu}\\
{\small Colorado State University}
}
\date{\today}
\maketitle

\begin{abstract}
We derive an estimator of the spectral density of a functional time series that
is the output of  a multilayer perceptron neural network. The estimator is motivated
by difficulties with the computation of existing  spectral density estimators
for time series
of functions defined on very large grids that arise, for example,  in climate 
compute models and medical scans. 
Existing estimators use autocovariance
kernels represented as  large $G\times G$ matrices, where $G$ is the
number of grid points on which the functions are evaluated. In many recent
applications,  functions are defined on 2D and 3D domains, and $G$ can be of
the order $G\sim 10^5$, making the  evaluation of the autocovariance
kernels computationally intensive or even impossible.
We use  the theory of spectral functional principal components to derive our
deep learning estimator and prove that it is a universal approximator
to the spectral density under general assumptions. 
Our estimator can be trained without
computing the autocovariance kernels and it  can be parallelized
to provide the estimates much faster than existing approaches.  
We validate its performance by simulations and an application to fMRI images.

\smallskip



\end{abstract}

\section{Introduction} \label{s:i}
Research on applications and improvements of
deep learning  has exploded in volume.
The overwhelming majority of contributions focus on new applications,
architectures, modes of training and similar issues that can be best
resolved by experimentation and extensive numerical studies. General
mathematical
foundations of deep learning were developed already in the 1990s,
including various
universal approximation and algorithm convergence results,
but there are still relatively few contribution studying deep learning in the framework
of mathematical statistics.
We propose a deep learning estimator
of the spectral density of functional time series and provide a
justification for its application under general assumptions.

Scalar second order stationary time series are described  by the mean and all
autocovariances, unlike an iid sample whose second order  parameters
are the mean and the variance.
In the case of scalar stationary observations $X_t$,
the spectral density is the Fourier transform of the autocovariance
sequence, i.e. it is defined by
\[
f(\theta) = \frac{1}{2\pi} \sum_{h=-\infty}^\infty c_h e^{-\mi h\theta},
\ \ \ \theta \in [-\pi, \pi],
\]
where $c_h = \cov(X_t, X_{t+h})$. The spectral density
contains all information about the second order dependence structure
because the autocovariances can be obtained by the inverse
Fourier transform.

A key motivation for this research is that high resolution  fMRI scans
are produced on 3D grids with $10^4$-$10^5$ points, so even estimating
covariances becomes difficult,  as it requires operations on
$10^8$-$10^{10}$ pairs of points. Our deep network  is trained directly on
the grid values, no sample autocovariances are required.
Computer climate models
produce functional time series on very large spatial domains, e.g. the
continent of North America and the surrounding ocean. For instance,  the
average  temperature
surface can be modeled at monthly resolution at a grid with a few kilometers
resolution, corresponding to about $10^5$ points.

This paper makes a contribution at the nexus
of deep learning and the analysis of functional time series.
There have been an increasing number of
papers on deep learning based inference, mostly  for  iid samples of functions.
We review them later in this section.
Our chief contribution is the derivation of neural networks that
estimate the spectral density of a  functional time series.
We justify the application of such networks under weak conditions
on the decay of autocovariance operators through several universal
approximation results in metrics relevant to the context we consider.
Our approach is based
on the frequency domain principal components
analysis of functional time series developed independently
by \citetext{panaretos:tavakoli:2013SPA}
and \citetext{hormann:dynamic:2015}. Estimation approaches presented
in those papers require computation of the autocovariance functions
at many lags, which may not be feasible if the domain on which
the functions are defined
consists of hundreds of thousands of dense grid points.
Any separable $L^2(\cQ)$ space  is isomorphic to
$L^2([0,1])$, but in practice a very large domain $\cQ$ makes the
estimation of even the covariance kernel computationally challenging,
as explained in \citetext{Panaretos:covnet}.
The data we consider are functions on $\cQ$ whose
values are observed on a dense grid. In applications that motivate
this work, there may be tens or hundreds of thousands of grid points,
so the estimation of the spectral density based on weighted sums of
autocovariances is computationally challenging
or even not feasible at present.
We show how to overcome this difficulty.
A chief contribution of our paper is to show how to combine the frequency
domain principal components analysis of functional time series
with deep learning by constructing suitable
output layers.  In numerical work, we use specific architectures for the deep layers,
those proposed by \citetext{Panaretos:covnet},
but they  could be modified in many ways without affecting our
theory. We develop a theoretical framework based on linear filters and
Fourier transforms of networks to show that our method is applicable
under very general assumptions.
In addition to \citetext{panaretos:tavakoli:2013SPA},
\citetext{hormann:dynamic:2015} and \citetext{Panaretos:covnet},
other closely related papers are
\citetext{panaretos:tavakoli:2013AS}, who derive a
mathematical framework for spectral analysis of functional
time series, and
\citetext{kartsioukas:SH:2023}  who
focus on  the estimation of the spectral density of a continuous domain
stationary process in a Hilbert space.
\citetext{kartsioukas:SH:2023}
obtain convergence rates and limit distributions for data observed on a grid,
which is the setting we consider in our numerical work.

We conclude this section with a brief review of recent work on the
application of deep learning to Functional Data Analysis. Our goal is
to give a general idea rather than list all important contributions.
\citetext{wang:ZCH:2024} provide an informative  review.
In most current  applications, functions are converted to
vectors by means of basis expansions.  Expansion coefficients form
vectors that can be used as inputs to a learning network.
Vector outputs can be converted back into functions,  or used
directly for other purposes, like classification or clustering.
Some useful  advancements to this approach have been made.
\citetext{yao:mueller:wang:2021} show how to construct and
imbed in a larger architecture and micro neural network that
learns a basis adaptively to the task. The context is of
scalar responses $y_i$ depending on functional regressors
$x_i(u)$, $u \in [0,1]$. Training is done on a sample of
iid realizations $(x_i, y_i)$.
\citetext{thind:multani:cao:2023} study a model with
multiple functional and scalar covariates.
Other contributions to advancing functional regression by application
of deep learning methods include
Rao and Reimherr
(\citeyear{rao:reimherr:2023sf}, \citeyear{rao:reimherr:2023ff}) and
\citetext{wu:beaulac:cao:2023}.
\citetext{hong:yao:mueller:wang:2024} consider the problem of
reconstructing latent trajectories $x_i(u)$, $u \in [0,1]$, from
noisy, sparsely observed realization $x(u_{ij}) + \eg_{ij}$,
$j=1, \ldots, n_i$, $i=1, \ldots, n$. They develop  GeLU-activated
transformers with augmented modules, basically custom-designed,
additional output layers that produce differentiable functions $x_i$.
Representation of functional data, including multivariate
functions, is considered by \citetext{wu:beaulac:cao:2024} and \citetext{wang:cao:2024}.
The above papers consider iid functions defined on a compact interval with
about $10^2$ grid points.
Regarding applications to time series of functions,
\citetext{wang:cao:2023} introduce an output
layer that improves predictions and
present applications to predicting air quality, electricity
price and mortality curves.  \citetext{ma:yao:zhou:2024} consider time series of
functions in a  context of traffic flow prediction.

The paper is organized as follows. Section \ref{sec:prel}
introduces the setting of functional time series and their spectral analysis.
In Section \ref{sec:spec},  we derive functional multilayer perceptrons suitable
for approximating the spectral density and formulate general universal approximation
results. We use the theory of Section \ref{sec:spec} to derive an
estimation algorithm in Section \ref{sec:impl}. Section
\ref{s:sim} contains a simulation study, while Section \ref{s:app} an
application to fMRI brain scans. The Supplementary Material
contains proofs and additional simulation results.

\section{Preliminaries }\label{sec:prel}
We introduce in this section the framework in which the methodology
and theory we propose operate, and fix the relevant notation.

Recall that $\cQ$ is  a compact subset of $\mathbb{R}^d$ and
define $ L^2(\cQ)$ to be the Hilbert space of
square integrable complex-valued functions on $\cQ$
equipped with the inner product
$\langle f, g \rangle = \int_\cQ f(u)\overline{g}(u) du$
and the induced norm $\Vert f \Vert = \sqrt{\langle f, f \rangle}$.
Here, $\overline{\cdot
}$ denotes the complex conjugate.
 For  $f,g \in L^2(\cQ)$, the operator $f \otimes g$ is defined by $(f \otimes g) (h) = \langle h,g\rangle f$. The operator $f \otimes g$ is a kernel operator
with the kernel $f\otimes g (u,v) = f(u) \overline{g}(v)$
i.e.  $(f \otimes g) (h)(u) = \int_\cQ  f(u) \overline{g}(v) h(v) dv$. We   use $f \otimes g$ to indicate both the operator and its kernel.
Note that $\Vert  f \otimes g\Vert_{\mathcal{S}} = \Vert  f \otimes g (\cdot,\cdot) \Vert_{L^2(\cQ\times \cQ)} $,
where  the left hand side
denotes the Hilbert-Schmidt norm of the operator and the right hand side
 denotes the $L^2(\cQ\times \cQ)$ norm of its kernel. The Hilbert-Schmidt inner product is denoted by $\langle \cdot, \cdot \rangle_{\mathcal{S}}$. Integral Hilbert-Schmidt operators on $L^2(\cQ)$ can be identified with their kernels in $L^2(\cQ\times \cQ)$. Therefore, for brevity, we may use $\Vert \cdot\Vert_{\mathcal{S}}$ instead of $\Vert \cdot\Vert_{L^2(\cQ\times \cQ)}$ when referring to the kernels. Detailed exposition of the theory of operators in Hilbert
 spaces is provided in \citetext{hsing:eubank:2015}.

Suppose $\{X_t\}_{t \in \mathbb{Z}}$ is an $L^2(\cQ)$-valued weakly stationary process i.e. $\mathbb{E} X_t = \mu$ and  $\mathbb{E} [ (X_{t+h} - \mu) \otimes (X_t - \mu) ]= C_h$, for all $t,h \in \mathbb{Z}$.
In most real data scenarios, the random fields are real-valued, and this is the assumption we make. However, since we are working in the frequency domain, employing a complex vector space will be advantageous.
Assumption \ref{as:sum:Ch} below  guarantees the existence of the
spectral density operator.

\begin{assumption}\label{as:sum:Ch}
The  process $\{X_t\}_{t \in \mathbb{Z}}$ is
weakly stationary in the space
$L^2(\cQ)$, has mean zero, and its autocovariance operators satisfy
$\sum_{h \in \mathbb{Z}} \Vert C_h \Vert_{\mathcal{S}} < \infty$.
(We assume that $X_t(u)$ is real for each $u\in \cQ$.)
\end{assumption}

In practice, we center the data by subtracting the sample mean.
It is  well-known  that the sample mean converges to the true mean
at the  rate of $\sqrt{N}$ ($N$ is the sample size)
under quite general assumptions,
see e.g. \citetext{horvath:kokoszka:reeder:2013},
so its estimation has an asymptotically negligible effect,
and is not considered here so as not to distract from the main contribution.

Under Assumption \ref{as:sum:Ch},
we define the spectral density operator $F^{X}(\theta)$ at the frequency $\theta \in [- \pi , \pi]$ by
\begin{align*}
    F^{X}(\theta) = \frac{1}{2 \pi } \sum_{h \in \mathbb{Z}} C_h \exp (-\mi h \theta),
\end{align*}
where $\mi = \sqrt{-1}$ is the imaginary unit and the convergence holds in the Hilbert-Schmidt norm. For each $\theta \in [- \pi , \pi]$, the spectral density operator $F^{X}(\theta)$ is a non-negative definite, Hilbert-Schmidt, self-adjoint operator.
For each $h$, the cross covariance operator $C_h$ is an integral  operator with kernel $c_h (u,v) = \mathbb{E}[X_h(u) X_0(v)    ]$,  i.e. $C_h(f)(u) = \int_\cQ c_h (u,v) f(v) dv$, for all $f \in L^2(\cQ)$.
This implies that  for each $\theta \in [- \pi , \pi]$
the spectral density operator $ F^{X}(\theta)$ is an integral operator with the kernel
\begin{align}\label{e:fX}
    f^{X}(\theta) (u,v) = \frac{1}{2 \pi }
    \sum_{h \in \mathbb{Z}} c_h (u,v) \exp (-\mi h \theta), \quad u,v \in \cQ,
\end{align}
where the convergence holds in $L^2(\cQ \times \cQ)$.
As in the coherence  analysis, we can write
\begin{align*}
 f^{X}(\theta) (u,v) =  p^{X}(\theta) (u,v) - \mi  q^{X}(\theta) (u,v),
\end{align*}
where
\begin{align}\label{eq:cos}
   p^{X}(\theta) (u,v) =  \frac{1}{2 \pi } \sum_{h \in \mathbb{Z}} c_h (u,v) \cos (h \theta)
\end{align}
is the cospectrum and
\begin{align}\label{eq:sin}
   q^{X}(\theta) (u,v) =  \frac{1}{2 \pi } \sum_{h \in \mathbb{Z}} c_h (u,v) \sin (h \theta)
\end{align}
is the quadspectrum. Note that under our assumption that $X_t(u)$ is real,
both  the cospectrum and quadspectrum are real-valued functions.
The estimation of the complex valued kernel $ f^{X}$
thus reduces to the estimation of two real-valued kernels.

We now present key results of  the frequency domain principal components
analysis of functional time series.
Since for each $\theta\in [-\pi, \pi]$,  $F^{X}(\theta)$ is  non-negative definite, Hilbert-Schmidt and self-adjoint, we have the spectral decomposition
\begin{align}\label{eq:s:dec}
    f^{X}(\theta) (u,v) =  \sum_{m \geq 1} \lambda_m(\theta) \varphi^{\dagger}_m(\theta)(u)\overline{\varphi}^{\dagger}_m(\theta)(v), \quad u,v \in \cQ,
\end{align}
with  nonnegative eigenvalues $\lambda_m(\theta)$ and the eigenfunctions $\varphi^{\dagger}_m(\theta)$   that form an orthonormal set in $L^2(\cQ)$. The pairs
$(\lambda_m(\theta), \varphi^{\dagger}_m(\theta))$ are arranged in the  decreasing order of eigenvalues.

Consider the functions
\begin{align*}
    \varphi_{m,h}(u) = \frac{1}{2\pi} \int_{-\pi}^{\pi} \exp(-\mi h \theta) \varphi^{\dagger}_m(\theta)(u)d \theta, \quad u \in \cQ,
     \ \ \ h\in \mbZ.
\end{align*}
Then, we have
\begin{align}\label{e:L}
\underset{L \rightarrow \infty}{\lim} \int_{-\pi}^{\pi}  \left\Vert \sum_{h= -L}^L \exp(\mi h \theta)\varphi_{m,h}- \varphi^{\dagger}_m(\theta)\right\Vert^2_{L^2(\cQ)}  d \theta = 0, \quad m \geq 1,
\end{align}
see Subsection 3.3 in \citetext{hormann:dynamic:2015}. The random field $\{X_t\}$ can be retrieved via
\begin{align}\label{eq:x:ret}
    X_t  = \sum_{m \geq 1} \sum_{h \in \mathbb{Z}} Y_{m, t+h} \varphi_{m,h},
\end{align}
where $Y_{m, t} =  \sum_{h \in \mathbb{Z}} \langle X_{t-h},\varphi_{m,h} \rangle$, and the convergence holds in mean square. The sequences $\{Y_{m, t}\}$ and $\{Y_{m', t}\}$ are uncorrelated at all lags if $m \neq m'$.

Our approach uses an approximation analogous to \refeq{L}
with appropriately constructed deep networks. The following
definition therefore plays a key role.

\begin{definition} (Fourier transformability) \label{def:FT}
We call a sequence $\{g_h\}_h \subset L^2(\cQ)$ {\em Fourier transformable} if
there exists a family $\{g^{\dagger}(\theta)\}_{\theta} \subset L^2(\cQ)$
such that
\begin{align}\label{eq:FT}
\lim_{L \to \infty} \int_{-\pi}^{\pi}
\left\Vert \sum_{h= -L}^L \exp (\mi h \theta)g_{h}- g^{\dagger}(\theta)
\right\Vert^2_{L^2(\cQ)}  d \theta = 0.
\end{align}
\end{definition}
Since $g^{\dagger}(\theta) \in L^2(\cQ)$,
$
\int_{-\pi}^{\pi} \left\Vert g^{\dagger}(\theta)\right\Vert^2_{L^2(\cQ)} d \theta
= \int_{-\pi}^{\pi} \int_\cQ |g^{\dagger}(\theta)(u) |^2 du d\theta
$
is the squared norm in $L^2( [-\pi, \pi] \times \cQ)$, a complete space, we conclude
that  $\int_{-\pi}^{\pi}
\left\Vert  g^{\dagger}(\theta)\right\Vert^2_{L^2(\cQ)}  d \theta < \infty$.

We will work with the class $\mathcal{A}$ of functions
of $m$ and $\theta$ defined as
\begin{equation}\label{e:cA}
\mathcal{A} =
\{\eta_{\cdot}(\cdot): \mathbb{D} \times [-\pi , \pi] \rightarrow
[0, \infty), \quad {\rm for \ some}  \   \mathbb{D} \subseteq \mathbb{N},\;
\underset{m,\theta}{\sup}\; \eta_m(\theta) < \infty\}.
\end{equation}
Equivalently,
$
  \mathcal{A} = \bigcup_{\mathbb{D} \subseteq \mathbb{N}}  \mathcal{A}_ \mathbb{D},
$
where
$
 \mathcal{A}_ \mathbb{D} =
\{\eta_{\cdot}(\cdot): \mathbb{D} \times [-\pi , \pi] \rightarrow
[0, \infty), \sup_{m,\theta}\eta_m(\theta) < \infty\}$,
$\mathbb{D} \subseteq \mathbb{N}$.
Each function in $\mathcal{A}$ is nonnegative  and
they are all bounded   above. The following lemma
is a direct consequence of Proposition 7 in \citetext{hormann:dynamic:2015}.

\begin{lemma}\label{lem:sum:lbd}
Suppose  Assumption \ref{as:sum:Ch} holds.
Then, for each $m$, the function $\theta \mapsto \lambda_m(\theta)$
is continuous. In particular, $\lambda_{\cdot}(\cdot) \in
\mathcal{A}_{\mathbb{N}} \subset \mathcal{A}$  because   $ \Lambda^{\ast} : =  \underset{m,\theta}{\sup}\; \lambda_m(\theta) = \underset{\theta}{\sup} \; \lambda_1(\theta) < \infty$.
\end{lemma}

We conclude this section with a list of functions, along with their domains and
ranges, that are frequently used throughout the paper.
The functions in the bottom three rows are introduced in
Section \ref{sec:spec}. We use the fraktur font to indicate networks.
\begin{align*}
f^X: \ & [-\pi , \pi] \rightarrow L^2 (\cQ \times \cQ),
    \quad \text{ or equivalently} \quad
F^X : \ [-\pi , \pi] \rightarrow \mathcal{S}; \\
    \mathfrak{f}: \ & [-\pi , \pi] \rightarrow L^2 (\cQ \times \cQ),
    \quad \text{ or equivalently}  \quad \mathfrak{f}: [-\pi , \pi]
    \rightarrow \mathcal{S}; \\
\varphi^{\dagger},
\mathfrak{g}^{\ddagger}: \ & [-\pi , \pi]\rightarrow  \ L^2(\cQ);   \\
  \varphi , \mathfrak{g}:& \mathcal{Q \rightarrow \mathbb{C}} \quad \text{  such that } \varphi ,
   \mathfrak{g} \in L^2(\cQ).
\end{align*}
We use $f^X$ to denote the kernel and $F^X$ to denote the corresponding
operator. In contrast, we use the same notation $\mathfrak{f}$
for both the operator and its kernel.
We recall that the space $L^2(\cQ)$ consists of
complex-valued functions.

\section{Spectral density approximation with deep networks}\label{sec:spec}
In this section,  we explain the mathematical mechanism for
approximating  of the spectral density operators with deep networks.
We do it through  theorems similar in spirit to
universal approximation results for neural networks. We first introduce
shallow networks  that will be ingredients of the output layer.
The deep layers form standard multilayer perceptrons, possibly with shared
parameters.
The building block networks we consider are similar to those introduced
in \citetext{Panaretos:covnet};  the key advance is in showing how to
transform and combine them to construct approximations to spectral density operators.

Consider the  following class of complex-valued shallow neural
networks defined on $\cQ$:
\begin{align*}
   \mathcal{C}^{\text{\tiny sh}} =  &  \{  \mathfrak{g} (\cdot ) = \sum_{r=1}^{R}
{c_r}
   \sigma (w_{r}^{\top} \cdot + b_{r}),  \quad R \in \mathbb{N},  w_{r} \in \mathbb{R}^d,  b_{r} \in \mathbb{R},  {c_r \in \mathbb{C}}\}\\
   =: & \{  \mathfrak{g} = \sum_{r=1}^{R}
g_{r}  ,  \quad R \in \mathbb{N}  \}.
    \end{align*}

The activation function $\sg: \mbR \to \mbR$ is always applied elementwise.
Given the hyperparameter $R \in \mathbb{N}$ and
$\sigma(\cdot)$, the parameters of these networks that
must be learned are
$w_{r} \in \mathbb{R}^d,  b_{r} \in \mathbb{R}, c_{r} \in \mathbb{C}$.

We next  introduce the class of deep shared neural networks. For positive integers $J, d_1, d_2, \ldots, d_J$, matrices $ W_1 \in \mathbb{R}^{d_1 \times d},  W_2 \in \mathbb{R}^{d_2 \times d_1},\ldots,  W_J \in \mathbb{R}^{d_J \times d_{J-1}}$, vectors $ B_1 \in \mathbb{R}^{d_1}, \ldots , B_J \in \mathbb{R}^{d_J}, w_r \in \mathbb{R}^{d_J} $,  scalars $b_r \in \mathbb{R}$, and $ {c_r \in \mathbb{C}}$, define
\begin{align}
\nonumber
  u_1  =& \sigma (W_1 u + B_1), \quad u \in \cQ,\\ \nonumber
u_{j+1}   =& \sigma (W_{j+1} u_j + B_{j+1}), \quad j=1,2,\ldots , J-1, \\  \label{eq:g:ds}
  g_r (u)  =&   {c_r} \sigma (w_{r}^{\top} u_J + b_{r}),\quad r=1,\ldots R.
\end{align}
In this architecture, we have neural networks $g_r(\cdot)$ with depth $J$ and width $\max \{d_1,\ldots d_J\}$. We assume that the first $J-1$ layers  are shared among $g_r(\cdot)$, $r=1,\ldots R$, and  only the last layer varies with $r$. This defines the following class of deep shared neural networks defined on $\cQ$:
\begin{align*}
   \mathcal{C}^{\text{\tiny ds}}
   = & \{  \mathfrak{g}  = \sum_{r=1}^{R}
g_{r}  ,  \quad R \in \mathbb{N}, \;\; \text{the }  g_{r} \text{ are of the form of \eqref{eq:g:ds}} \}.
    \end{align*}

Relaxing the assumption of shared weights and biases,
we define general deep neural networks
\begin{align}
\nonumber
  u_{1,r}  =& \sigma (W_{1,r} u + B_{1,r}), \quad u \in \cQ,\\ \nonumber
u_{j+1,r}   =& \sigma (W_{j+1,r} u_{j,r} + B_{j+1,r}), \quad j=1,2,\ldots , J-1, \\  \label{eq:g:d}
  g_r (u)  =&  {c_r} \sigma (w_{r}^{\top} u_{J,r} + b_{r}),\quad r=1,\ldots R.
\end{align}
In this construction, in addition to allowing the parameters of the last layer to vary, we also permit all matrix and vector parameters of the other hidden layers to vary with $r$.
This defines  the following class of deep  neural networks defined on $\cQ$:
\begin{align*}
   \mathcal{C}^{\text{\tiny d}}
   = & \{  \mathfrak{g}   = \sum_{r=1}^{R}
g_{r}  ,  \quad R \in \mathbb{N}, \;\; \text{the } g_{r}  \text{ are  in the form of \eqref{eq:g:d}} \}.
    \end{align*}

For the sake of brevity, we use the notation $\mathcal{C}^{\text{\tiny nn}}$ to indicate any of the classes $\mathcal{C}^{\text{\tiny sh}}$, $\mathcal{C}^{\text{\tiny ds}}$, or $\mathcal{C}^{\text{\tiny d}}$.  A generic element of $\mathcal{C}^{\text{\tiny nn}}$  is denoted by $
\mathfrak{g}  $. We emphasize that each element of $\mathcal{C}^{\text{\tiny nn}}$ is a  function in $L^2(\cQ)$ of a specific form known as a multilayer neural network.

We now  define the class of
sequences of  such neural networks that are eventually zero:
 \begin{align}\label{eq:cNN}
    \mathcal{C} = \{\{\mathfrak{g}_h  \}_{h\in \mathbb{Z}}:  \; \mathfrak{g}_h
    \in \mathcal{C}^{\text{\tiny nn}} \text{ if } |h| \leq L, \
    \mathfrak{g}_h = 0 \text{ if } |h| > L,
    \text{ for some  } L \in \mathbb{N} \},
\end{align}
that is, $\mathcal{C}$ encompasses the sequences of complex-valued neural networks
with  finitely many non-zero elements. In particular, these sequences
satisfy Definition \ref{def:FT}, i.e. are trivially Fourier transformable.
In the following, we use the notation $\mathfrak{g}^{\ddagger}$ rather than
$\mathfrak{g}^{\dagger}$ to emphasize that no limit is needed
in the case of networks. We thus
define the class $\mathcal{D}$ containing the  Fourier transforms
of the sequences  in  $\mathcal{C}$:
\begin{align}\label{eq:cFNN}
    \mathcal{D} = \{ \mathfrak{g}^{\ddagger}: [-\pi , \pi]
    \rightarrow L^2(\cQ),\; \mathfrak{g}^{\ddagger}:  \theta  \mapsto \sum_{h \in \mathbb{Z}} \exp(\mi h \theta) \mathfrak{g}_h , \;  \text{for some  }\{\mathfrak{g}_h \}\in \mathcal{C}  
    \}.
\end{align}
Finally, recall the class $\mathcal{A}$ is defined in \refeq{cA},  and
define the class $\mathcal{E}$:
\begin{align}\label{eq:cF}
    \mathcal{E}
 = & \{ \mathfrak{f}: [-\pi , \pi] \rightarrow \mathcal{S}, \ \mathfrak{f}: \theta \mapsto \sum_{m=1}^M
     \eta_{m} (\theta)
     \mathfrak{g}^{\ddagger}_{m}( \theta) \otimes \mathfrak{g}^{\ddagger}_{m} (\theta), \; \\ \nonumber
     & \;\;\; \text{for some  }  M \in \mathbb{N},     \eta_{\cdot} (\cdot) \in \mathcal{A},  \;\mathfrak{g}^{\ddagger}_{m}  \in  \mathcal{D}, m=1, \ldots M\}.
\end{align}
Note that if $\mathfrak{f} \in \mathcal{E}$, then for each $\theta \in [-\pi, \pi]$, $\mathfrak{f}(\theta)$ is a nonnegative and self-adjoint operator,
just like the spectral density  $F^X(\theta)$ at the frequency $\theta$. We will show in Section \ref{sec:impl} that the networks  $\mathfrak{f}(\theta)$ can be trained without the need to estimate the autocovariances $c_h(u,v)$ appearing in \eqref{e:fX}.
Theorem \ref{thm:uni} below states that every spectral density can  be approximated in the integrated Hilbert-Schmidt norm  by neural networks in $\mathcal{E}$ under the following general assumption.

\begin{assumption}\label{as:sg}
    The activation function $\sigma(\cdot)$ is such that for any
    $\epsilon >0$ and any $\varphi \in L^2(\cQ)$ there is
    a network $\mathfrak{g}$ in $\mathcal{C}^{\text{\tiny nn}}$
    such that $\Vert \varphi -\mathfrak{g}\Vert_{L^2(\cQ)} < \epsilon$.
\end{assumption}

We verify in Section \ref{s:ua} of the Supplementary material that
all practically used activation functions satisfy
Assumption \ref{as:sg}.

\begin{theorem}\label{thm:uni}
Suppose  Assumptions \ref{as:sum:Ch} and \ref{as:sg} hold.
Then, for any $\epsilon >0$,  there exists $\mathfrak{f} \in \mathcal{E}$   such that
\begin{align*}
  \int_{-\pi}^{\pi} \Vert f^{X}(\theta)  - \mathfrak{f}(\theta) \Vert_{\mathcal{S}} d \theta < \epsilon.
\end{align*}
\end{theorem}
Theorem \ref{thm:uni}  cannot be used directly to construct a deep learning estimator.
We therefore modify the universal approximation formulated in Theorem \ref{thm:uni}
and  state a similar result in terms of the Fourier transform of a sequence of networks.
This paves the way for the construction of the estimators in Section \ref{sec:impl}.
In light of \eqref{eq:x:ret}, for $M , L\in \mathbb{N}$,
consider  the stationary  sequence of random fields
\begin{align}\label{eq:tfX}
 \widetilde{\mathfrak{X}}_t = \sum_{h =-L}^{L}
 \sum_{m=1}^M \xi_{m,t+h} \mathfrak{g}_{m, h},
\end{align}
where, for each pair $(m,h)$, $\mathfrak{g}_{m, h}  \in \mathcal{C}^{\text{\tiny nn}}$   and $\{\xi_t = \left( \xi_{1,t}, \ldots , \xi_{M,t}\right)\}$ is an $M$-dimensional mean zero  stationary random process with uncorrelated components at all lags. In particular, the spectral density operators  $F^{\xi}(\theta)$ are diagonal $M \times M$ matrices with non-negative entries.   Theorem \ref{thm:E:fX}  below states
that the elements of  the class $\mathcal{E}$, defined in \eqref{eq:cF},
can be viewed as spectral density kernels of the stationary  sequences
defined in \eqref{eq:tfX}.
Sequences of the form \eqref{eq:tfX} can be viewed as networks
with an  additional output layer parameterized by the $\xi_{m,t+h}$.
For ease of reference,  we formulate the following assumption.

\begin{assumption} \label{a:sta}
The neural random fields $ \widetilde{\mathfrak{X}}_t$ are in the form \eqref{eq:tfX}  for some $ M, L \in \mathbb{N}$, where, for each pair $(m,h)$, $\mathfrak{g}_{m, h}  \in \mathcal{C}^{\text{\tiny nn}}$  and the sequence
$\{\xi_t = \left( \xi_{1,t}, \ldots , \xi_{M,t}\right)\}$ is an $M$-dimensional
mean zero  stationary random process with uncorrelated components
at all lags and absolutely summable autocovariance matrices. In particular, the
long-run variance  matrix  $F^{\xi}$ is a diagonal $M \times M$
matrix with non-negative entries.
\end{assumption}

\begin{theorem}\label{thm:E:fX}
Let $ \{\widetilde{\mathfrak{X}}_t \}$
be a  sequence of random fields   satisfying  Assumption \ref{a:sta}.
Then,  $ \{\widetilde{\mathfrak{X}}_t \}$   is stationary
and
its spectral density kernel  has the representation
\begin{align}\label{eq:F:fX}
    F^{\widetilde{\mathfrak{X}}}(\theta) = & \sum_{m=1}^M
    F_{m,m}^{\xi} (\theta)
     \mathfrak{g}^{\ddagger}_{m}(\theta)     \otimes \mathfrak{g}_{m}^{\ddagger}(\theta),
\end{align}
where $F_{m,m}^{\xi}(\theta)$ is the $(m,m)$ entry  of the diagonal matrix $F^{\xi}(\theta)$ and $ \mathfrak{g}^{\ddagger}_{m}$ is the Fourier transform of the finite series $\{\mathfrak{g}_{m, h} \}_{-L \leq h \leq L} \in \mathcal{C}$, for $m=1,\ldots,M$. Moreover, the class $\mathcal{E}$ admits the representation
\begin{align}\label{e:alt-f}
\mathcal{E} =   \{ \{f^{\widetilde{\mathfrak{X}}}(\theta)\}_{\theta \in [-\pi , \pi]}, \ \widetilde{\mathfrak{X}}_t = \sum_{h = -L}^L \sum_{m=1}^M \xi_{m,t+h}
\mathfrak{g}_{m, h}, \text{as in Assumption \ref{a:sta}} \}.
\end{align}
\end{theorem}

\begin{remark}\label{r:real}
Since the random fields $X_t$ are real-valued,  each eigenfunction
$\varphi^{\dagger}_m(\theta)$ is Hermitian, i.e.
$\varphi^{\dagger}_m(\theta) = \overline{\varphi}^{\dagger}_m(-\theta)$.
This implies $\varphi_{m,h} = \overline{\varphi}_{m,h}$,
which in turn implies that the scores $Y_{m,t}$
appearing in \eqref{eq:x:ret} are real.
The networks $\mathfrak{g}_{m, h}  $ defined in \eqref{eq:tfX}
are therefore real-valued.
This  restricts the  class $\mathcal{D}$
to the Hermitian functions and the $\xi_{m,t}$ to real numbers.
\end{remark}

We now  address approximation of  the spectral
density kernel $f^X$ with  finite weighted sums of the autocovariances
of the network fields $\widetilde{\mathfrak{X}}_t$.
We consider three specific kernels,
and need to tighten the autocovariance summability
condition in Assumption \ref{as:sum:Ch} for some of these kernels.
Abstract assumptions could  be formulated, but it is  useful
to have specific assumptions that can be readily applied and make the
proofs more transparent.

We consider three commonly used kernels: the Truncated kernel
\begin{align*}
       \omega(s) = \left\{
       \begin{array}{cc}
           1, & \vert s \vert \leq 1 ,\\
           0, & \vert s \vert > 1,
       \end{array}
       \right.
\end{align*}
the Bartlett kernel
\begin{align*}
    \omega(s) = \left\{
    \begin{array}{cc}
      1- \vert s \vert,   &  0 \leq \vert s \vert \leq 1,\\
        0,  & \vert s \vert >1,
    \end{array}
    \right.
\end{align*}
and the Parzen kernel
\begin{align*}
       \omega(s) = \left\{
       \begin{array}{cc}
       1-6\vert s \vert^2 +   6\vert s \vert^3,  & \vert s \vert < \frac{1}{2}, \\
         2 (1-\vert s \vert )^3,   & \frac{1}{2} \leq \vert s \vert \leq 1, \\
         0, & \vert s \vert > 1.
       \end{array}
       \right.
\end{align*}

We work under the following Assumption.

\begin{assumption} \label{a:3k} Suppose $\og(\cdot)$ is either
the Truncated, the Bartlett or the Parzen Kernel. If either the
Bartlett or the Parzen Kernel is used, assume  that
for some $0 < \alpha \leq 1$, $\sum_{h \in \mathbb{Z}} \vert h \vert^{\alpha}
\left \Vert C^X_h \right \Vert_{\mathcal{S}} < \infty$. Assume that
$q \to \infty$ and $q/N\to 0$,  as $N \to \infty$.
\end{assumption}

\begin{theorem}\label{thm:wq}
Let $\{X_t\}_{t \in \mathbb{Z}}$
be a  stationary processes satisfying  Assumption
\ref{as:sum:Ch}.
Suppose also that   Assumptions \ref{as:sg} and  \ref{a:3k} hold.
Then, for any $\epsilon >0$, there is  $q \geq 1$
and networks $\{{\mathfrak{X}}_t\}$
satisfying Assumption \ref{a:sta}  such that
    \begin{align*}
         \int_{-\pi}^{\pi} \lnorm f^{X}(\theta)  -
         \sum_{|h|\leq q}     \omega\lp\frac{h}{q}\rp C^{{\mathfrak{X}}}_h
         \exp (-\mi h \theta) \rnorm_{\mathcal{S}} d \theta < \epsilon.
    \end{align*}
\end{theorem}

\begin{remark}\label{rem:L2}
Inspection of the proof of Theorem \ref{thm:wq} shows that
other approximation results could be derived: the squared norm could be
used and/or
$\sum_{|h|\leq q} \omega\lp h/q \rp C^{X}_h \exp (-\mi h \theta)$
in place of $ f^{X}(\theta)$.
We do not list those variants to conserve space.
\end{remark}

\section{Construction of network estimators}\label{sec:impl}
Suppose we observe a  realization  $\{X_1, \ldots, X_N\}$.
For each $t =1, \ldots, N$, we
approximate $X_t$ by a network $\widetilde{\mathfrak{X}}_t$
given by \eqref{eq:tfX} with coefficients $\xi_{m,t} \in \mathbb{R}$
and networks $\mathfrak{g}_{m, h}  \in \mathcal{C}^{\text{\tiny nn}}$
that must be learned. The $\xi_{m,t}$ are treated in this section
as unknown parameters, not random sequences.

In light of Theorems \ref{thm:uni} and \ref{thm:wq} and Remark
\ref{rem:L2}, we choose the loss function
\begin{equation}\label{eq:loss}
  \ell := \int_{-\pi}^{\pi} \lnorm \hat{f}^{X}(\theta)
  - \hat{f}^{\widetilde{\mathfrak{X}}}(\theta)\rnorm_{\mathcal{S}} d \theta
  = \int_{-\pi}^{\pi} \lnorm \sum_{|h|\leq q}
  \omega\lp\frac{h}{q}\rp
  \lb \widehat{C}^{X}_h - \widehat{C}^{{\mathfrak{X}}}_h \rb
  \exp (-\mi h \theta)
  \rnorm_{\mathcal{S}} d \theta.
\end{equation}
In this loss function,  $\hat{f}^{X}$ and $\hat{f}^{\widetilde{\mathfrak{X}}}$
are estimators of the spectral densities  ${f}^{X}$ and $f^{\widetilde{\mathfrak{X}}}$, respectively, such  that the optimization problem is linear in the count, say $G$,  of grid points in $\cQ$ at which the fields $X_t$ are observed. Calculations presented in this section will show how to construct such
a training algorithm.

Consider the empirical autocovariance operators
\[
\widehat{C}^X_h = \frac{1}{N} \sum_{k=1}^{N-h} X_{h+k} \otimes X_k,
\quad  h \geq 0;  \qquad\qquad \widehat{C}^X_h =  \frac{1}{N}
\sum_{k=1}^{N-|h|} X_{k} \otimes X_{|h|+k}, \quad h <0.
\]
and  the lag window estimators  based on the observed
fields $X_1, \ldots, X_N$ by
\begin{equation}\label{eq:Fhat}
\widehat{F}^{X}(\theta) =  \frac{1}{2 \pi } \sum_{|h|\leq q} \omega\lp \frac{h}{q}\rp
\widehat{C}^X_h \exp(-\mi h \theta)
=    \sum_{ |h| \leq q} \tilde\omega \lp \frac{h}{q}\rp
\left(\frac{1}{N} \sum_{k=1}^{N} X_{h+k} \otimes X_k \right),
\end{equation}
setting the terms with implausible subscripts to zero and
\[
\tilde\omega(h,\theta) := \frac{1}{2 \pi } \omega(h/q) \exp (-\mi h \theta).
\]
Likewise, for the approximating network fields
$\{\widetilde{\mathfrak{X}}_t \}_{t=1}^N$  in \eqref{eq:tfX}, we define
\begin{equation}\label{eq:pFhat}
\widehat{F}^{\widetilde{\mathfrak{X}}}  (\theta)
    =  \sum_{ |h| \leq q} \tilde\omega(h,\theta)
    \widehat{C}^{\widetilde{\mathfrak{X}}}_h,
\end{equation}
where
\[
\widehat{C}^{\widetilde{\mathfrak{X}}}_h =
    \frac{1}{N} \sum_{k=1}^{N} \widetilde{\mathfrak{X}}_{h+k}
    \otimes \widetilde{\mathfrak{X}}_k
    = \frac{1}{N} \sum_{k=1}^{N}  \sum_{j,j'=-L}^L
    \sum_{m,m'=1}^M \left( \xi_{m,h+k+j} \xi_{m',k+j'} \right)\mathfrak{g}_{m, j} \otimes \mathfrak{g}_{m', j'},
\]
setting  the terms with implausible subscripts to zero.

As noted in Section \ref{sec:prel}, there is a one-to-one correspondence
between the operators $\widehat{F}^{X}(\theta)$ and
$\widehat{F}^{\widetilde{\mathfrak{X}}}  (\theta)$
and  their respective kernels $\hat{f}^{X}(\theta)$, $\hat{f}^{\widetilde{\mathfrak{X}}}(\theta)$.
The evaluation of these kernels requires the number of operation
proportional to $G^2$, so they will not be used directly in the
training algorithm. However, the squared Hilbert-Schmidt norm of
the difference can be written as a sum of four terms each of which
can be evaluated using $O(G)$ operations.  To see this,   observe that

\begin{align*}
 \lnorm  \widehat{F}^{X}(\theta)  - \widehat{F}^{\widetilde{\mathfrak{X}}}(\theta) \rnorm^2_{\mathcal{S}} & =  \frac{1}{N^2} \lip  \sum_{ h=-q}^q \sum_{k=1}^{N}
   \tilde\omega(h,\theta)
      \left( X_{h+k} \otimes X_k - \widetilde{\mathfrak{X}}_{h+k} \otimes \widetilde{\mathfrak{X}}_k \right), \right.\\
      &\ \ \ \ \ \ \ \ \ \ \left. \sum_{ h=-q}^q
      \sum_{k=1}^{N}\tilde\omega(h,\theta)
      \left( X_{h+k} \otimes X_k - \widetilde{\mathfrak{X}}_{h+k} \otimes \widetilde{\mathfrak{X}}_k \right)\rip_{\mathcal{S}}\\  \nonumber
& =  \frac{1}{N^2}\sum_{ h=-q}^q
      \sum_{k=1}^{N}   \sum_{ h'=-q}^q \sum_{k'=1}^{N}
      \tilde\omega(h,\theta)  \overline{\tilde\omega(h',\theta)}
      \\
 & \ \ \ \      \lip
      \left( X_{h+k} \otimes X_k - \widetilde{\mathfrak{X}}_{h+k} \otimes \widetilde{\mathfrak{X}}_k \right) ,
      \left( X_{h'+k'} \otimes X_{k'} - \widetilde{\mathfrak{X}}_{h'+k'} \otimes \widetilde{\mathfrak{X}}_{k'} \right)
      \rip_{\mathcal{S}}.
 \end{align*}
Therefore,
\begin{align}\label{eq:|F-F|2}
&\lnorm  \widehat{F}^{X}(\theta)  - \widehat{F}^{\widetilde{\mathfrak{X}}}(\theta) \rnorm^2_{\mathcal{S}}\\       &=  \frac{1}{N^2} \sum_{ h=-q}^q
      \sum_{k=1}^{N}   \sum_{ h'=-q}^q \sum_{k'=1}^{N}
      \tilde\omega(h,\theta)  \overline{\tilde\omega(h',\theta)}
      \langle  X_{h+k},  X_{h'+k'}\rangle  \langle  X_{k},  X_{k'}\rangle \nonumber \\
      & \ \ -\frac{1}{N^2} \sum_{ h=-q}^q \sum_{k=1}^{N}   \sum_{ h'=-q}^q \sum_{k'=1}^{N}
      \tilde\omega(h,\theta)  \overline{\tilde\omega(h',\theta)}
      \langle X_{h+k}, \widetilde{\mathfrak{X}}_{h'+k'}\rangle  \langle X_{k}, \widetilde{\mathfrak{X}}_{k'} \rangle  \nonumber \\
      &\ \ -\frac{1}{N^2} \sum_{ h=-q}^q
      \sum_{k=1}^{N}   \sum_{ h'=-q}^q \sum_{k'=1}^{N}
      \tilde\omega(h,\theta)  \overline{\tilde\omega(h',\theta)}
      \langle \widetilde{\mathfrak{X}}_{h+k}, X_{h'+k'}\rangle  \langle \widetilde{\mathfrak{X}}_{k}, X_{k'}\rangle   \nonumber\\
      &\ \  + \frac{1}{N^2}\sum_{ h=-q}^q
      \sum_{k=1}^{N}   \sum_{ h'=-q}^q \sum_{k'=1}^{N}
      \tilde\omega(h,\theta)  \overline{\tilde\omega(h',\theta)}
      \langle \widetilde{\mathfrak{X}}_{h+k},\widetilde{\mathfrak{X}}_{h'+k'}\rangle  \langle \widetilde{\mathfrak{X}}_{k}, \widetilde{\mathfrak{X}}_{k'}\rangle. \nn
\end{align}

Note that
\[
\tilde\omega(h,\theta)  \overline{\tilde\omega(h',\theta)} = \frac{1}{4 \pi^2 }
 \og\lp \frac{h}{q}\rp \og\lp \frac{h'}{q}\rp
\lbr \cos\lp [h-h']\theta\rp  - \mi  \sin\lp [h-h']\theta\rp \rbr.
\]
By Remark \ref{r:real}, the $\widetilde{\mathfrak{X}}_{t}$ are real,
so in an optimization algorithm,  the  products
$\tilde\omega(h,\theta)  \overline{\tilde\omega(h',\theta)}$
on the right-hand side of \eqref{eq:|F-F|2} can be replaced
with  real numbers
\[
r(h, h'; \theta) = \frac{1}{4 \pi^2 }
\og\lp \frac{h}{q}\rp \og\lp \frac{h'}{q}\rp
\cos\lp [h-h']\theta\rp.
\]

Representation \eqref{eq:|F-F|2} shows that the computation of
$\Vert \widehat{F}^{X}(\theta)  -
\widehat{F}^{\widetilde{\mathfrak{X}}}(\theta) \Vert^2_{\mathcal{S}}$
involves only expressions linear in the count $G$ of the grid points in $\cQ$ at which the fields $X_t$ are observed, avoiding computations quadratic in $G$. This also applies to its square root  $\Vert \widehat{F}^{X}(\theta)  - \widehat{F}^{\widetilde{\mathfrak{X}}}(\theta) \Vert_{\mathcal{S}}$, which appears in the loss function \eqref{eq:loss}, and   has the key impact on  the computational feasibility of the spectral density estimation problem for time series of random fields defined on large
domains. These calculations lead to the following algorithm.

\medskip
\noindent  {\sc Algorithm 1} (Spectral density estimation):

\medskip
\noindent \textbf{  Step 1:} Construct $\{\widetilde{\mathfrak{X}}_t\}_{t=1}^N$ according to \eqref{eq:tfX},
with real $\xi_{m,h}$ and real valued networks  $\mathfrak{g}_{m,h}$.
Given the hyperparameters  $M$,  $L$ and $q$,  the parameter vector is
\[\vartheta = \{ \xi_{m,h} \text{ and the parameters of  } \mathfrak{g}_{m,h}, \; 1 \leq m \leq M, \; \vert h \vert \leq L+q\}.
\]

\medskip
\noindent  \textbf{  Step 2:}
Use the constructed $\{\widetilde{\mathfrak{X}}_t\}_{t=1}^N$ and the observed $\{X_t\}_{t=1}^N$ to compute  \eqref{eq:|F-F|2} using a   discrete grid on  $ [-\pi,\pi]$ and $\cQ$.

\medskip
\noindent  \textbf{   Step 3:}
Compute the numerical integral, over $\theta$,  of the square root of \eqref{eq:|F-F|2}. This produces a numerical version of the loss function $\ell$ defined in \eqref{eq:loss}, which we denote by   $\hat{\ell}$.

\medskip
\noindent  \textbf{   Step 4:} Minimize $\hat{\ell}$ over the parameters specified  in  Step 1. Call this minimizer $\hat\vartheta$.

 \medskip
\noindent  \textbf{   Step 5:} Plug in the minimizer $\hat\vartheta$
to obtain
\[
\widehat{C}^{\widetilde{\mathfrak{X}}}_h (\hat\vartheta)
=   \frac{1}{N} \sum_{k=1}^{N}  \sum_{j,j'=-L}^L \sum_{m,m'=1}^M \left( \hat\xi_{m,h+k+j} \hat\xi_{m',k+j'} \right)
\hat{\mathfrak{g}}_{m, j} \otimes \hat{\mathfrak{g}}_{m', j'},
\]
where that hats over the $\xi$s and the networks indicate their evaluations at the optimized values.

\medskip
\noindent  \textbf{   Step 6:} Compute the estimated  cospectrum
\begin{align*}
   \hat{p}^{X}(\theta) (u,v) =  \frac{1}{2 \pi }  \sum_{|h| \le q}
   \omega\lp\frac{h}{q}\rp
\hat{c}^{\widetilde{\mathfrak{X}}}_h (\hat\vartheta)(u,v)
\cos (h \theta)
\end{align*}
and the quadspectrum
\begin{align*}
   \hat{q}^{X}(\theta) (u,v) =  \frac{1}{2 \pi }   \sum_{|h| \le q}
   \omega\lp\frac{h}{q}\rp
\hat{c}^{\widetilde{\mathfrak{X}}}_h(u,v)   \sin (h \theta).
\end{align*}
(The coordinates $(u,v)$ appear only in the trained networks
$\hat{\mathfrak{g}}_{m, j}$.)

\medskip

Step 4 is the learning process of the feedforward networks we consider.
Details, including the selection of the hyperparameters,  are discussed in
Section \ref{s:sim}. We refer to the estimator defined by Algorithm 1 as
the {\em spectral-NN estimator}.

\section{Numerical implementation and  simulations}\label{s:sim}
We have shown in previous sections that the spectral-NN estimator
is a universal approximator of  the spectral density of a  functional time
series under assumptions  on
the network and the kernel/bandwidth that practically
always hold, Assumptions \ref{as:sg} and \ref{a:3k}.
In this section, we compare
via simulations the spectral-NN estimator to the lag-window
estimator \eqref{eq:Fhat}
that has been studied and used in previous work, e.g.
\citetext{hormann:dynamic:2015} and \citetext{kuenzer2021principal}.
We will see that the spectral-NN estimator is basically never worse,
often much better, and if the count of grid points is very
large, it is the only estimator that can actually be computed.
As with all simulation studies, such conclusion cannot be established
with the generality of mathematical results, but they provide useful
insights. The code for implementation of our method are available
at \url{https://github.com/sohamsarkar1991/spectral-NN}.

We consider the simplest functional time series model,
$X_t = \gamma X_{t-1} + Z_t$,  $\gamma \in (-1,1)$.
We first generate discrete observations  of a white noise (innovation) process
$Z_t = \{Z_t (u),\; u \in  [0,1]^d\}$, $t=1,2,\ldots, N$.
This is fully discussed in \citetext{Panaretos:covnet},
who generate independent  Gaussian random fields  on a grid
for $d=2,3$.  The distribution
of the initial value $X_0$ is taken to be the same as $Z_1$.
To ensure approximate stationarity, we  generate a time series
of length $N+N_0$, and discard the first $N_0$ elements
(we use $N_0=100$ in our simulations).

We also need to obtain the closed form of the spectral density operators
$F^X(\theta)$  for the purpose of comparing them to the estimated objects.
For the innovation process $\{Z_t\}$,
$F^Z(\theta) = \frac{1}{2 \pi}C_0^Z$, $\theta \in [-\pi , \pi]$,
where $C_0^Z$ is the lag-$0$ covariance operator of  $\{Z_t\}$.
The causal representation of the process $\{X_t\}$ implies
\begin{align*}
    F^X(\theta)
 =& \left(\sum_{h=0}^{\infty} \gamma^{h} \exp(-\mi h \theta)\right) \frac{1}{2 \pi}C_0^Z \left(\sum_{h=0}^{\infty} \gamma^{h} \exp(\mi h \theta)\right)\\
    = & \left(1-\gamma \exp(-\mi \theta)\right)^{-1} \frac{1}{2 \pi}C_0^Z  \left(1-\gamma \exp(\mi \theta)\right)^{-1}\\
    =& \frac{1}{2 \pi} \frac{C_0^Z}{1+\gamma^2-2\gamma \cos (\theta)}.
\end{align*}
For the lag-$0$ covariance operator $C_0^Z$
(equivalently, the covariance kernel $c_0^Z$)
we make three choices similar to \citetext{Panaretos:covnet}, viz.
\begin{itemize}
	\item[(i)] \emph{Brownian sheet}: $c_0^Z(u,v) = \min\{u_1,v_1\} \times \cdots \times \min\{u_d,v_d\}, \, u,v \in [0,1]^d$. For $d=1$, this reduces to the standard Brownian motion.
	\item[(ii)] \emph{Integrated Brownian sheet}: $c_0^Z(u,v) = c_{\rm ibm}(u_1,v_1) \times \cdots \times c_{\rm ibm}(u_d,v_d), \, u,v \in [0,1]^d$, where $c_{\rm ibm}$ is the covariance kernel of the integrated Brownian motion, defined as $c_{\rm ibm}(u,v) = \int_{0}^u \int_{0}^v \min\{s,t\} ds dt$.
	\item[(iii)] \emph{Mat\'ern}: $c_0^Z(u,v) = 2^{1-\nu}/\Gamma(\nu)\, (\sqrt{2\nu}\,\|u-v\|_d)^\nu\,K_{\nu}(\sqrt{2\nu}\,\|u-v\|_d),\, u,v \in [0,1]^d$, where $\Gamma$ is the gamma function, $K_{\nu}$ is the modified Bessel function of the second kind and $\|\cdot\|_d$ is the Euclidean distance on $\mathbb{R}^d$. The Mat\'ern covariance model is indexed by the smoothness parameter $\nu > 0$. We use $\nu=0.001,0.01,0.1$ and $1$ in our simulation studies.
\end{itemize}
These covariance models produce a wide variety of smoothness structures on the generated random fields. Particularly, the random fields generated using the Brownian sheet are continuous but nowhere differentiable, whereas they are continuously differentiable for the integrated Brownian sheet. For the Mat\'ern covariance, larger values of $\nu$ results in smoother random fields; see \citetext{Panaretos:covnet} for details.

We consider simulations with $d=1,2$ and $3$, which we refer to as 1D, 2D and 3D, respectively. The random fields are generated on a $K \times \cdots \times K$
regular grid on $[0,1]^d$.
We need to choose the sample size $N$, the grid size parameter  $K$,
and the autoregression coefficient $\gamma$.
We consider three different setups by fixing two of these parameters
and varying the third one. To evaluate the performance of the spectral-NN estimator,
we consider the relative estimation error
\[
\frac{\int_{-\pi}^{\pi} \|F^{X}(\theta) - {\widehat F}^{\widetilde{\mathfrak X}}(\theta)\|_{\mathcal S} d\theta}{\int_{-\pi}^{\pi} \|F^{X}(\theta)\|_{\mathcal S} d\theta},
\]
where ${\widehat F}^{\widetilde{\mathfrak X}}(\cdot)$ is the estimated spectral density using the neural network model.
Since, the integrals cannot be computed in closed forms, we use Monte-Carlo approximation to the integrals. In particular, we generate a random sample $\theta_1,\ldots,\theta_I$ from the uniform distribution on $[-\pi,\pi]$, and for each $i=1,\ldots, I$, we generate a random sample $(u_{i1},v_{i1}),\ldots,(u_{iJ},v_{iJ})$ from the uniform distribution on $[0,1]^d \times [0,1]^d$, to approximate
\begin{align*}
\frac{1}{2\pi} \int_{-\pi}^{\pi} \|F^{X}(\theta) - {\widehat F}^{\widetilde{\mathfrak X}}(\theta)\|_{\mathcal S} d\theta &\approx \frac{1}{I}\sum_{i=1}^I \left[\frac{1}{J} \sum_{j=1}^{J} \left\{f^{X}(\theta_i)(u_{ij}, v_{ij}) - {\widehat f}^{\widetilde{\mathfrak X}}(\theta_i)(u_{ij}, v_{ij})\right\}^2 \right]^{1/2}, \\
\frac{1}{2\pi} \int_{-\pi}^{\pi} \|F^{X}(\theta)\|_{\mathcal S} d\theta &\approx \frac{1}{I}\sum_{i=1}^I \left[\frac{1}{J} \sum_{j=1}^{J} \left\{f^{X}(\theta_i)(u_{ij}, v_{ij})\right\}^2 \right]^{1/2}.
\end{align*}
Finally, the ratio of these two quantities gives an approximation to the relative error.
In our simulations, we used $I=100$ and $J=10000$. We proceed analogously
to compute the relative estimation error of the lag-window estimator \eqref{eq:Fhat}.
In the tables that follow, we refer to these two estimators, respectively,
as NN and Emp.

In all our simulations, we use the deep spectral-NN estimator. We also need to select a few hyperparameters: $M$, $L$, the depths, widths and activation functions of the neural networks
$\mathfrak g_{m,h}$. For both estimators, we need to select
the truncation level $q$ and the weight kernel $\omega$.
Throughout our simulations, we use the \emph{sigmoid}
$\sigma(t) = (1+e^{-t})^{-1}$ as the activation function.
We ran a few pilot simulations with different choices of
$M$ ($5,10,20$), $L$ ($5,10,20$), depth ($2,3,4,5,6$),
width ($10,20,30,40,50$), $q$ ($5,10,20,40$) and $\omega$
(truncated, Bartlett, Parzen, Tukey-Hanning, quadratic spectral).
In those simulations, we observed that the results were not affected
much by the choice of these hyperparameters, except for $q$ and $\omega$.
The results with $q=20,40$ were much better than those with $q=5,10$.
However, larger values of $M,L$ and $q$ result in higher computing times.
Based on our experience, we use $M=L=10$, depth=$4$, width=$20$,
and $q=20$ throughout our simulation study. For the weight function,
the Parzen kernel produced the best results in our pilot study, which we use throughout.
In order to have a fair comparison,
we use $q=20$ and the Parzen kernel for the empirical (lag-window) estimator as well.

\begin{table}[t]
\centering
\footnotesize
\caption{Relative error rates (in \%) of the empirical spectral density estimator (Emp) and the spectral-NN estimator (NN) in different 2D examples. The results are for a fixed AR coefficient $\gamma=0.5$, fixed resolution $K=50$, and varying sample size $N$. The numbers are averages based on $25$ simulation runs. The corresponding standard errors are in the next line in italics and in a smaller font.\label{tab:2D_N}}
\bigskip

\setlength{\tabcolsep}{0.04in}
\begin{tabular}{r c rr c rr c rr c rr c rr c rr}
&& \multicolumn{2}{c}{} && \multicolumn{2}{c}{Integrated} && \multicolumn{11}{c}{} \\
&& \multicolumn{2}{c}{Brownian} && \multicolumn{2}{c}{Brownian} && \multicolumn{11}{c}{Matern} \\
&& \multicolumn{2}{c}{Sheet} && \multicolumn{2}{c}{Sheet} && \multicolumn{2}{c}{$\nu=0.001$} && \multicolumn{2}{c}{$\nu=0.01$} && \multicolumn{2}{c}{$\nu=0.1$} && \multicolumn{2}{c}{$\nu=1$} \\
\cline{3-4} \cline{6-7} \cline{9-10} \cline{12-13} \cline{15-16} \cline{18-19}
$N$ && Emp & NN && Emp & NN && Emp & NN && Emp & NN && Emp & NN && Emp & NN \\ [4pt]
$100$	&& $47.95$	& $43.32$	&& $32.44$	& $34.10$	&& $1966.41$	& $52.66$	&& $1766.47$	& $56.67$	&& $209.21$	& $67.71$	&& $43.32$	& $42.38$\\
	&& $\se{1.74}$	& $\se{1.80}$	&& $\se{1.82}$	& $\se{1.77}$	&& $\se{4.41}$	& $\se{2.72}$	&& $\se{4.43}$	& $\se{1.61}$	&& $\se{1.36}$	& $\se{1.60}$	&& $\se{1.16}$	& $\se{1.17}$\\ [2pt]
$200$	&& $33.07$	& $30.01$	&& $23.77$	& $24.74$	&& $1416.19$	& $46.68$	&& $1279.16$	& $52.39$	&& $151.35$	& $52.46$	&& $30.86$	& $30.58$\\
	&& $\se{0.70}$	& $\se{0.68}$	&& $\se{1.10}$	& $\se{1.04}$	&& $\se{1.66}$	& $\se{1.28}$	&& $\se{2.04}$	& $\se{1.62}$	&& $\se{0.47}$	& $\se{0.69}$	&& $\se{0.94}$	& $\se{0.96}$\\ [2pt]
$400$	&& $23.76$	& $22.38$	&& $16.13$	& $16.53$	&& $1016.09$	& $43.25$	&& $916.12$	& $47.97$	&& $108.31$	& $39.07$	&& $23.56$	& $23.36$\\
	&& $\se{0.59}$	& $\se{0.79}$	&& $\se{0.89}$	& $\se{0.94}$	&& $\se{1.19}$	& $\se{1.04}$	&& $\se{1.19}$	& $\se{1.01}$	&& $\se{0.27}$	& $\se{0.47}$	&& $\se{0.79}$	& $\se{0.84}$\\ [2pt]
$800$	&& $17.78$	& $17.41$	&& $11.84$	& $12.69$	&& $730.87$	& $42.32$	&& $659.79$	& $47.70$	&& $77.95$	& $31.72$	&& $16.45$	& $16.52$\\
	&& $\se{0.52}$	& $\se{0.54}$	&& $\se{0.64}$	& $\se{0.83}$	&& $\se{0.61}$	& $\se{0.77}$	&& $\se{0.56}$	& $\se{0.99}$	&& $\se{0.16}$	& $\se{0.33}$	&& $\se{0.56}$	& $\se{0.56}$\\ [2pt]
$1600$	&& $13.13$	& $13.86$	&& $8.44$	& $9.83$	&& $531.17$	& $41.83$	&& $479.56$	& $45.87$	&& $56.32$	& $26.59$	&& $11.49$	& $11.98$\\
	&& $\se{0.45}$	& $\se{0.44}$	&& $\se{0.43}$	& $\se{0.63}$	&& $\se{0.47}$	& $\se{0.67}$	&& $\se{0.41}$	& $\se{0.58}$	&& $\se{0.17}$	& $\se{0.28}$	&& $\se{0.45}$	& $\se{0.45}$
\end{tabular}
\end{table}

We repeat each simulation setup $25$ times and report the
average relative errors along with the corresponding standard errors.
In Tables~\ref{tab:2D_N}--\ref{tab:2D_gamma},
we report these results for the 2D setup only. The results for the 1D and 3D
setups are similar, which are reported in the supplementary material
(see Appendix~\ref{sec:additional_simulations}).

\begin{table}[h!]
\centering
\footnotesize
\caption{Relative error rates (in \%) of the empirical spectral density estimator (Emp) and the spectral-NN estimator (NN) in different 2D examples. The results are for a fixed AR coefficient $\gamma=0.5$, fixed sample size $N=250$, and varying resolution $K$. The numbers are averages based on $25$ simulation runs. The corresponding standard errors are in the next line in italics and in a smaller font. A dash (---) indicates that the program failed due to insufficient memeory.\label{tab:2D_K}}
\bigskip

\setlength{\tabcolsep}{0.04in}
\begin{tabular}{r c rr c rr c rr c rr c rr c rr}
&& \multicolumn{2}{c}{} && \multicolumn{2}{c}{Integrated} && \multicolumn{11}{c}{} \\
&& \multicolumn{2}{c}{Brownian} && \multicolumn{2}{c}{Brownian} && \multicolumn{11}{c}{Matern} \\
&& \multicolumn{2}{c}{Sheet} && \multicolumn{2}{c}{Sheet} && \multicolumn{2}{c}{$\nu=0.001$} && \multicolumn{2}{c}{$\nu=0.01$} && \multicolumn{2}{c}{$\nu=0.1$} && \multicolumn{2}{c}{$\nu=1$} \\
\cline{3-4} \cline{6-7} \cline{9-10} \cline{12-13} \cline{15-16} \cline{18-19}
$K$ && Emp & NN && Emp & NN && Emp & NN && Emp & NN && Emp & NN && Emp & NN \\ [4pt]
$10$	&& $33.53$	& $31.58$	&& $26.47$	& $22.51$	&& $1413.66$	& $172.78$	&& $1276.75$	& $222.37$	&& $147.18$	& $82.57$	&& $29.47$	& $29.15$\\
	&& $\se{0.97}$	& $\se{0.95}$	&& $\se{0.99}$	& $\se{1.00}$	&& $\se{4.15}$	& $\se{6.59}$	&& $\se{4.01}$	& $\se{7.93}$	&& $\se{1.04}$	& $\se{1.59}$	&& $\se{1.06}$	& $\se{0.97}$\\ [2pt]
$20$	&& $31.19$	& $29.06$	&& $21.60$	& $20.53$	&& $1301.69$	& $80.31$	&& $1176.46$	& $84.45$	&& $138.13$	& $57.98$	&& $28.47$	& $28.08$\\
	&& $\se{0.76}$	& $\se{0.84}$	&& $\se{0.97}$	& $\se{1.06}$	&& $\se{2.83}$	& $\se{3.20}$	&& $\se{2.61}$	& $\se{3.02}$	&& $\se{0.78}$	& $\se{0.90}$	&& $\se{0.81}$	& $\se{0.82}$\\ [2pt]
$40$	&& $29.11$	& $26.60$	&& $18.28$	& $18.40$	&& $1275.89$	& $47.07$	&& $1153.01$	& $51.27$	&& $135.11$	& $47.21$	&& $26.70$	& $26.29$\\
	&& $\se{0.76}$	& $\se{0.79}$	&& $\se{0.98}$	& $\se{1.09}$	&& $\se{1.64}$	& $\se{1.39}$	&& $\se{1.53}$	& $\se{1.21}$	&& $\se{0.53}$	& $\se{0.76}$	&& $\se{0.81}$	& $\se{0.84}$\\ [2pt]
$80$	&& $30.72$	& $28.34$	&& $20.07$	& $20.26$	&& $1267.63$	& $42.38$	&& $1145.32$	& $47.85$	&& $135.58$	& $45.96$	&& $28.30$	& $27.85$\\
	&& $\se{0.80}$	& $\se{0.87}$	&& $\se{0.96}$	& $\se{1.10}$	&& $\se{1.12}$	& $\se{1.17}$	&& $\se{0.99}$	& $\se{0.94}$	&& $\se{0.54}$	& $\se{0.75}$	&& $\se{0.88}$	& $\se{0.87}$\\ [2pt]
$160$	&& ---	& $26.76$	&& ---	& $19.27$	&& ---	& $37.58$	&& ---	& $43.96$	&& ---	& $43.53$	&& ---	& $26.04$\\
	&& & $\se{0.79}$	&& & $\se{1.12}$	&&	& $\se{0.64}$	&& & $\se{0.53}$	&& & $\se{0.60}$	&&	& $\se{0.78}$
\end{tabular}
\end{table}

In Table~\ref{tab:2D_N}, the results are reported for $K=50$, $\gamma=0.5$ and different values of $N$. The results in this setup are the usual, the errors of both the estimators decrease as $N$ increases. For the coarser Brownian sheet example, the spectral-NN estimator performs slightly better than the empirical estimator. The scenario is reversed in the smoother integrated Brownian sheet example. The situation is remarkably different in the case of Mat\'ern covariance, particularly with smaller values of $\nu$, which corresponds to rougher surfaces.
In these examples, the empirical estimator fails to capture the underlying
spectral density. The spectral-NN estimator, on the other hand,
can successfully detect the underlying structure,
even from these rough observations.
Interestingly, the error of the empirical estimator can be $20$ times
(or even higher than) that of the spectral-NN estimator.

In Table~\ref{tab:2D_K}, we report the results with $N=250$, $\gamma=0.5$ and varying $K$. The results in this setup are qualitatively similar to the previous setup. For the Brownian sheet and integrated Brownian sheet, the errors are not affected by the resolution. However, for the Mat\'ern covariance, while the errors for the spectral-NN estimator decreases rapidly with the resolution, the same is not true for the empirical estimator. This again shows the adverse effects of roughness on the empirical estimator, which can be mitigated by using the proposed neural network structure.

\begin{table}[t]
\centering
\footnotesize
\caption{Relative error rates (in \%) of the empirical spectral density estimator (Emp) and the spectral-NN estimator (NN) in different 2D examples. The results are for a fixed sample size $N=250$, fixed resolution $K=50$, and varying AR coefficients $\gamma$. The numbers are averages based on $25$ simulation runs. The corresponding standard errors are in the next line in italics and in a smaller font.\label{tab:2D_gamma}}
\bigskip

\setlength{\tabcolsep}{0.04in}
\begin{tabular}{r c rr c rr c rr c rr c rr c rr}
&& \multicolumn{2}{c}{} && \multicolumn{2}{c}{Integrated} && \multicolumn{11}{c}{} \\
&& \multicolumn{2}{c}{Brownian} && \multicolumn{2}{c}{Brownian} && \multicolumn{11}{c}{Matern} \\
&& \multicolumn{2}{c}{Sheet} && \multicolumn{2}{c}{Sheet} && \multicolumn{2}{c}{$\nu=0.001$} && \multicolumn{2}{c}{$\nu=0.01$} && \multicolumn{2}{c}{$\nu=0.1$} && \multicolumn{2}{c}{$\nu=1$} \\
\cline{3-4} \cline{6-7} \cline{9-10} \cline{12-13} \cline{15-16} \cline{18-19}
$\gamma$ && Emp & NN && Emp & NN && Emp & NN && Emp & NN && Emp & NN && Emp & NN \\ [4pt]
$0.1$	&& $30.64$	& $28.24$	&& $18.24$	& $18.49$	&& $1269.82$	& $45.74$	&& $1147.81$	& $51.61$	&& $135.23$	& $44.88$	&& $27.82$	& $27.41$\\
	&& $\se{0.66}$	& $\se{0.72}$	&& $\se{0.71}$	& $\se{0.69}$	&& $\se{0.85}$	& $\se{1.01}$	&& $\se{1.09}$	& $\se{1.31}$	&& $\se{0.29}$	& $\se{0.51}$	&& $\se{0.65}$	& $\se{0.62}$\\ [2pt]
$0.25$	&& $30.15$	& $27.44$	&& $19.57$	& $19.49$	&& $1268.87$	& $46.81$	&& $1146.76$	& $50.32$	&& $135.50$	& $46.16$	&& $28.35$	& $28.10$\\
	&& $\se{0.71}$	& $\se{0.71}$	&& $\se{1.01}$	& $\se{0.91}$	&& $\se{1.07}$	& $\se{1.67}$	&& $\se{1.01}$	& $\se{1.28}$	&& $\se{0.46}$	& $\se{0.56}$	&& $\se{0.96}$	& $\se{1.02}$\\ [2pt]
$0.5$	&& $30.38$	& $27.97$	&& $20.75$	& $20.90$	&& $1272.52$	& $45.03$	&& $1150.08$	& $52.82$	&& $136.49$	& $48.56$	&& $28.12$	& $27.81$\\
	&& $\se{0.84}$	& $\se{0.83}$	&& $\se{1.23}$	& $\se{1.21}$	&& $\se{1.87}$	& $\se{1.52}$	&& $\se{1.54}$	& $\se{1.47}$	&& $\se{0.60}$	& $\se{0.95}$	&& $\se{0.82}$	& $\se{0.78}$\\ [2pt]
$0.75$	&& $35.00$	& $32.82$	&& $23.36$	& $24.09$	&& $1301.26$	& $49.80$	&& $1176.66$	& $53.70$	&& $138.13$	& $52.46$	&& $33.49$	& $33.08$\\
	&& $\se{1.34}$	& $\se{1.40}$	&& $\se{1.72}$	& $\se{1.62}$	&& $\se{5.00}$	& $\se{1.91}$	&& $\se{5.65}$	& $\se{2.11}$	&& $\se{0.77}$	& $\se{0.77}$	&& $\se{1.41}$	& $\se{1.46}$\\ [2pt]
$0.9$	&& $55.16$	& $52.94$	&& $53.31$	& $54.33$	&& $1414.54$	& $63.75$	&& $1316.36$	& $79.09$	&& $163.79$	& $77.81$	&& $52.28$	& $52.12$\\
	&& $\se{2.30}$	& $\se{1.80}$	&& $\se{3.67}$	& $\se{3.47}$	&& $\se{34.76}$	& $\se{2.47}$	&& $\se{34.54}$	& $\se{3.58}$	&& $\se{5.35}$	& $\se{3.09}$	&& $\se{1.85}$	& $\se{1.91}$
\end{tabular}
\end{table}

The empirical estimator could not be computed due to insufficient memory
for a resolution of $160 \times 160$. This is a ramification of the fact that the empirical estimator is highly demanding in terms of memory, since it requires computing empirical autocovariances, which are large dimensional objects. In particular, for observations on a $K \times K$ grid, the empirical autocovariances are $K^4$-dimensional objects. Moreover, several such ($2q+1$, to be precise) autocovariances need to be computed and stored for the empirical estimator, which can become prohibitive even for moderate values of $K$.

\begin{table}[h!]
\centering
\footnotesize
\caption{Average computing times (in seconds) and maximum memory usage (in MB) of the empirical spectral density estimator (Emp) and the spectral-NN estimator (NN) in different 2D examples. For NN, computing times with GPU are shown in the next line in italics. The codes were run on a computer with 64 GiB RAM, AMD Ryzen 9 5900X (3.7 GHz) CPU, NVIDIA GeForce RTX 3090 GPU, and Ubuntu 24.04.2 LTS (64-bit) OS. A dash (---) indicates that the program failed due to insufficient memeory.\label{tab:2D_time_memory}}
\bigskip

\setlength{\tabcolsep}{0.05in}
\begin{tabular}{l c rr c rr c rr c rr c rr}
\multicolumn{16}{c}{Fixed resolution of 50 $\times$ 50 and variying sample sizes ($N$).} \\\hline
$N$ && \multicolumn{2}{c}{$100$} && \multicolumn{2}{c}{$200$} && \multicolumn{2}{c}{$400$} && \multicolumn{2}{c}{$800$} && \multicolumn{2}{c}{$1600$} \\
 &&	Emp & NN && Emp & NN && Emp & NN && Emp & NN && Emp & NN \\ [3pt]
Fit && $2.86$ & $288.07$ && $5.51$ & $370.31$ && $10.69$ & $483.51$ && $20.42$ & $773.32$ && $40.91$ & $2116.57$ \\
    && & $\gpu{250.64}$ && & $\gpu{271.51}$ && & $\gpu{318.35}$ && & $\gpu{403.86}$ && & $\gpu{626.89}$ \\
Eval  && $428.48$ & $105.83$ && $425.26$ & $104.99$ && $421.25$ & $104.87$ && $428.28$ & $101.25$ && $425.37$ & $102.28$ \\
      && & $\gpu{1.85}$ && & $\gpu{1.85}$ && & $\gpu{1.85}$ && & $\gpu{1.85}$ && & $\gpu{1.89}$ \\
Total && $431.34$ & $393.90$ && $430.77$ & $475.30$ && $431.94$ & $588.38$ && $448.70$ & $824.56$ && $466.28$ & $2218.85$ \\
      && & $\gpu{252.49}$ && & $\gpu{273.36}$ && & $\gpu{320.20}$ && & $\gpu{405.71}$ && & $\gpu{628.72}$ \\ [2pt]
Memory && $1184$ & $700$ && $1185$ & $679$ && $1185$ & $692$ && $1187$ & $709$ && $1188$ & $694$ \\ [3pt]

\multicolumn{16}{c}{Fixed sample size of 250 and variying resolutions.} \\\hline
$K$ && \multicolumn{2}{c}{$10$} && \multicolumn{2}{c}{$20$} && \multicolumn{2}{c}{$40$} && \multicolumn{2}{c}{$80$} && \multicolumn{2}{c}{$160$} \\
 &&	Emp & NN && Emp & NN && Emp & NN && Emp & NN && Emp & NN \\ [3pt]
Fit && $0.22$ & $165.71$ && $0.33$ & $198.36$ && $0.92$ & $296.76$ && $346.41$ & $746.18$ && --- & $2698.37$ \\
    && & $\gpu{278.12}$ && & $\gpu{279.04}$ && & $\gpu{278.55}$ && & $\gpu{288.59}$ && & $\gpu{315.84}$ \\
Eval  && $48.86$ & $103.24$ && $95.02$ & $98.18$ && $281.34$ & $98.27$ && $1092.12$ & $98.96$ && --- & $99.50$ \\
      && & $\gpu{1.86}$ && & $\gpu{1.85}$ && & $\gpu{1.86}$ && & $\gpu{1.86}$ && & $\gpu{1.86}$ \\
Total && $49.08$ & $268.96$ && $95.35$ & $296.54$ && $282.25$ & $395.03$ && $1438.53$ & $845.14$ && --- & $2797.87$ \\
      && & $\gpu{279.98}$ && & $\gpu{280.89}$ && & $\gpu{280.41}$ && & $\gpu{290.45}$ && & $\gpu{317.70}$ \\ [2pt]
Memory && $120$ & $696$ && $150$ & $696$ && $564$ & $700$ && $7143$ & $839$ && --- & $2224$
\end{tabular}
\end{table}

To further demonstrate this, in Table~\ref{tab:2D_time_memory}, we report the maximum memory requirements by the empirical estimator and the spectral-NN estimator for different 2D examples with varying values of $N$ and $K$ (on a computer with 64 GiB RAM, AMD Ryzen 9 5900X (3.7 GHz) CPU, NVIDIA GeForce RTX 3090 GPU, and Ubuntu 24.04.2 LTS 64-bit operating system). From the table, it can be observed that the maximum memory requirement of the empirical estimator increases exponentially with $K$. For $K=160$, the empirical estimator requires more than $64$ gigabytes of memory, compared to $2.2$ gigabytes for the spectral-NN estimator. The situation can be much worse in 3D, where the empirical estimator can fail even for moderate resolutions of $25 \times 25 \times 25$ (see Table~\ref{tab:3D_time_memory} in Appendix~\ref{sec:additional_simulations}).

In Table~\ref{tab:2D_time_memory}, we also report the average runtimes of the two estimators. For both the estimators, the computations have two components: fitting the model and evaluating the model for error computation. For the empirical estimator, the fitting includes computing and storing the autocovariances, while for the spectral-NN estimator this includes estimating the parameters of the model. The fitting part is relatively less time consuming for the empirical estimator, but the evaluation part is highly demanding, especially when $K$ is large. For the spectral-NN estimator, on the other hand, the fitting part can be quite time consuming. But once the model is fitted, evaluation is very fast. This shows the utility of the proposed estimator compared to the empirical estimator in terms of applicability. Moreover, the computing time of the spectral-NN estimator can be substantially lowered using GPU computing, especially when $K$ or $N$ (or both) is large. In fact, Table~\ref{tab:2D_time_memory} shows that we get almost $4$ times reduction in computing time for $N=1600$ and more than $9$ times reduction for $K=160$. The computing times can be further reduced by considering other modern machine learning techniques like mini-batch learning, although we did not implement it in our simulations.

In Table~3, we report the results for $N=250$, $K=50$ and different
values of the autoregression coefficient $\gamma$.
In this setup, the problem becomes harder when the value of $\gamma$ increases,
though the relative performance of the two estimators remain similar.

\section{Application to a time series of brain scans} \label{s:app}
To further demonstrate the usefulness of the spectral-NN estimator,
we use it on a 3D fMRI data. We consider brain scans of subject
\texttt{sub69518}
from Beijin from the \emph{1000 Functional Connectomes Project}
(\url{https://www.nitrc.org/projects/fcon_1000/}).
The data consist of 3D brain scans taken at a resolution
of $64 \times 64 \times 33$ over $225$ time points separated by $2$ seconds.
These data sets were previously analyzed by \citetext{aston2012}
and \citetext{stoehr2021} who concluded that they are stationary
after standard voxel-wise preprocessing.
\citetext{Panaretos:covnet} used their CovNet
method to estimate the covariance of these  data treating
them  as i.i.d.\ functional observations.
However, these data are actually a functional time series because
the scans separated by $2$ seconds are likely to be dependent.

Before applying the estimator, we pre-processed the data by removing the first $5$ time points. To mitigate the edge-effect, we also removed the first three and last three voxels from the $x$-axis and $y$-axis; and the first two and last two voxels from the $z$-axis. This gave us a time series of $220$ 3D scans at a resolution of $59 \times 59 \times 29$ each. As suggested by \citetext{aston2012,Panaretos:covnet}, we removed a polynomial trend of order $3$ from each voxel and scaled the data to have voxel-wise unit variance.

\begin{figure}[t]
\centering
\includegraphics[width=0.78\textwidth]{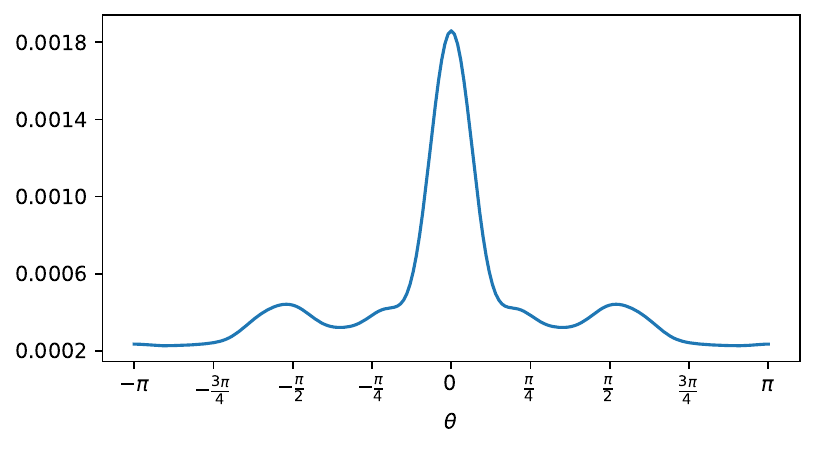}
\caption{The magnitude of the fitted spectral-NN estimator for the 3D fMRI data.
The spectral-NN model was fitted with $M=L=10$, depth$=4$, width$=20$
and $q=20$.\label{fig:fMRI_data}}
\end{figure}

In this example, the spectral-NN estimator
$\widehat{f}^{\widetilde{\mathfrak X}}$
is a function over $[-\pi,\pi] \times [0,1]^6$.
To visualize the estimator, we obtain its magnitude
at different frequencies $\theta \in [-\pi,\pi]$.
That is, we compute
$\|\widehat{F}^{\widetilde{\mathfrak X}}(\theta)\|_{\mathcal S}
= \sqrt{\langle \widehat f^{\widetilde{\mathfrak X}}(\theta),
\widehat f^{\widetilde{\mathfrak X}}(\theta)\rangle_{\mathcal S}}$
for $\theta \in [-\pi,\pi]$.
These magnitudes are shown in Figure~\ref{fig:fMRI_data}.
The figure shows that the magnitude of the estimated spectral
density varies with $\theta$. If the scans formed a functional
white noise, the curve in  Figure~\ref{fig:fMRI_data} would be (approximately)
a constant horizontal line.
This indicates that there is indeed temporal dependence in the data.
The general shape is consistent with an AR(1) model used in
Section \ref{s:sim}, but there is a bump at $\pi/2$. The graph
shows only the norms, but we can see that the spectral-NN estimator
is a promising tool for the analysis of functional time series on large
domains.

\medskip

\centerline{ \sc Acknowledgement}

This research was   partially supported by the United
States National Science Foundation grant  DMS--2412408.
The research of Soham Sarkar was partially supported by the INSPIRE Faculty
Fellowship from the Department of Science and Technology, Government of India.

\medskip

\centerline{ \sc Supplementary material}

The Supplementary Material contains proofs and additional simulation results.

\bigskip
\bibliographystyle{oxford3}
\renewcommand{\baselinestretch}{0.9} \small
\bibliography{nedaDL}

\newpage

\appendix

\centerline{\large SUPPLEMENTARY MATERIAL}

\bigskip

\section{Universal approximation in the space $L^2(\cQ)$}
\label{s:ua}
Commonly used activation functions are described,
for example, in Section 6.2.3 of \citetext{BB:2024}, and include
ReLU, leaky ReLU, hard tanh, tanh, softplus and logistic sigmoid.
They are all continuous functions, either piecewise linear with one or two
points where the derivative does not exist, or infinitely differentiable functions
that are not polynomials. We can therefore use the results of
\citetext{leshno:1993} to establish the following proposition.
Recall that $\cQ$ is a compact subset of $\mbR^d$.

\begin{proposition}\label{prop:npol}
If the activation function
$\sigma$ is not a polynomial, then
each class $\mathcal{C}^{\text{\tiny nn}}$ is dense
in $L^2(\cQ)$.
\end{proposition}
\noindent{\sc Proof.} Each class $\mathcal{C}^{\text{\tiny nn}}$ contains
the class $\mathcal{C}^{\text{\tiny sh}}$,  which coincides with the
class $\Sigma_d$ considered in Theorem 1 of \citetext{leshno:1993},
except that \citetext{leshno:1993} consider real-valued functions
and we consider complex-valued functions. Their results can be applied
to the real and imaginary parts. Proposition 1 of \citetext{leshno:1993}
then implies that $\mathcal{C}^{\text{\tiny sh}}$ is dense in
any space $L^p(\mu)$, $1 \le p < \infty$,  as long as $\mu$ is absolutely
continuous with respect to Lebesgue measure on $\mbR^d$. In particular,
$\mathcal{C}^{\text{\tiny sh}}$ is dense in $L^2(\cQ)$.
\hfill
\QED

\section{Preliminary lemmas}\label{p:prel}
For ease of reference we state here two  lemmas
frequently used in the proofs. The first lemma follows directly
from Theorem 2.1 in \citetext{kokoszka_frequency_2020},
the second from the fact that the $\varphi^{\dagger}_m(\theta)$
are orthonormal and from Lemma \ref{lem:Lbd}.

\begin{lemma} \label{lem:Lbd}
Suppose  Assumption \ref{as:sum:Ch}  holds. Then
    \begin{align*}
 \infty > \mathbb{E} \Vert X_0 \Vert^2  = \sum_{m \geq 1} \int_{-\pi}^{\pi} \lambda_m (\theta) d \theta =: \sum_{m \geq 1} \Lambda_m =:  \Lambda.
\end{align*}
\end{lemma}

\begin{lemma}\label{lem:M}
Set \[f_M^{X}(\theta) (u,v):= \sum_{m =1}^M \lambda_m(\theta) \varphi^{\dagger}_m(\theta)(u)\overline{\varphi}^{\dagger}_m(\theta)(v),  \quad u,v \in \cQ.\]
Under Assumption \ref{as:sum:Ch},     for any $\epsilon > 0$,   there exists $M$ such that
\begin{align*}
   \int_{-\pi}^{\pi} \Vert f^{X}(\theta)  - f_M^{X}(\theta) \Vert_{\mathcal{S}} d \theta < \epsilon.
\end{align*}
\end{lemma}

\section{Proofs of the results of Section  \ref{sec:spec}}\label{p:spec}
Before proceeding with the proofs, we review the
required background information.
For a more comprehensive discussion, we refer to  subsection 3.3 of \citetext{hormann:dynamic:2015}.
Recall the spectral density decomposition \eqref{eq:s:dec}, which we can write
as
\begin{equation} \label{e:FXdec}
  F^{X}(\theta) =  \sum_{m \geq 1} \lambda_m(\theta)
  \varphi^{\dagger}_m(\theta)\otimes \varphi^{\dagger}_m(\theta).
\end{equation}
Since for every $\theta$ the functions $\varphi^{\dagger}_m(\theta)$,
$m \ge 1$, are
orthonormal, $\| \varphi^{\dagger}_m(\theta)\| = 1$,  and so
\begin{align}\label{eq:ph^2}
\int_{\cQ}   \int_{-\pi}^{\pi}
\left\vert\varphi^{\dagger}_m(\theta)(u)\right\vert^2 d \theta  du
= \int_{-\pi}^{\pi} \int_{\cQ}
\left\vert\varphi^{\dagger}_m(\theta)(u)\right\vert^2 du d \theta = 2\pi < \infty.
\end{align}
Therefore, for almost all $u \in \cQ$,
$\int_{-\pi}^{\pi} \left\vert\varphi^{\dagger}_m(\theta)(u)\right\vert^2
d \theta < \infty$.
Denoting by $\mathcal{L}eb(\cdot)$ the Lebesgue measure on $\mbR^d$,
there are  thus $A_m\subseteq \cQ$, with
$\mathcal{L}eb(A_m) = \mathcal{L}eb(\cQ) < \infty$,
such that $\int_{-\pi}^{\pi}
\left\vert\varphi^{\dagger}_m(\theta)(u)\right\vert^2 d \theta < \infty$,
for all $u \in A_m$.
Define
\begin{equation} \label{eq:varphi}
\varphi_{m,l} (u) = \left \{
\begin{array}{ll}
\frac{1}{2\pi}\int_{-\pi}^{\pi}
\varphi^{\dagger}_m(\theta)(u) \exp (-\mi l \theta) d \theta, & u \in A_m, \\
0, & u \notin A_m.
\end{array}
\right.
\end{equation}
The sets $A_m$ are introduced only to have $\varphi_{m,l} (u)$ defined
at every $u\in \cQ$, they do  not affect any mean-square convergence
results. In particular,
the inversion formula \eqref{eq:x:ret} continues to hold
(in the mean square sense), i.e.
\begin{align}\label{eq:X:inv}
X_t = \sum_{m=1}^{\infty} \sum_{l \in \mathbb{Z}} Y_{m, t+l} \varphi_{m,l},
\ \ \ {\rm where} \ \  Y_{m, t} = \sum_{l \in \mathbb{Z}}
\langle X_{t-l}, \varphi_{m,l}\rangle.
\end{align}
Moreover, for each $m$, $\varphi^{\dagger}_m(\theta)$ and $\varphi_{m,h}$
are connected through Definition \ref{def:FT}.

\bigskip
\noindent
{\sc Proof of Theorem \ref{thm:uni} .}
Fix  $\epsilon > 0$.
Using the triangle  inequality, for  $\mathfrak{f} \in \mathcal{E}$
and a  positive integer $M$, we have
\begin{align}\label{eq:fX-f}
\nonumber
  &  \int_{-\pi}^{\pi}  \Vert f^{X}(\theta) - \mathfrak{f}(\theta)
  \Vert_{\mathcal{S}}  d \theta \\
  \leq  &
   \int_{-\pi}^{\pi} \Vert f^{X}(\theta)  -  f^{X}_{M}(\theta)
   \Vert_{\mathcal{S}} d \theta
    +  \int_{-\pi}^{\pi} \Vert f^{X}_{M}(\theta)
    - \mathfrak{f}(\theta) \Vert_{\mathcal{S}} d \theta,
\end{align}
where $f_M^{X}(\theta) $ is defined in Lemma \ref{lem:M},
which implies that there is  a sufficiently large $M$, such that
\begin{align}\label{eq:s1}
    \int_{-\pi}^{\pi} \Vert f^{X}(\theta) - f_{M}^{X}(\theta)
    \Vert_{\mathcal{S}}  d\theta  < \epsilon/2.
\end{align}
In the following, we fix this $M$ and focus on the second
term that involves a network approximation.
We will find  networks $\mathfrak{g}_{m,h}$, $m=1, 2, \ldots, M$,
such that $\{\mathfrak{g}_{m,h}\}_h \in \mathcal{C}$,  that make
the second term in \eqref{eq:fX-f} smaller than $\epsilon/2$.
 Recall that
$\varphi^{\dagger}_m(\theta)$ and $\varphi_{m,h}$
are connected through relation \eqref{eq:FT}.
For each $m=1, 2, \ldots, M$ and  positive integer $L$,
define $c = c(L) = 2^{L-1}$ and  choose the neural networks
$\mathfrak{g}_{m,h}(\cdot)$ such that
\begin{align}\label{eq:nn:prf}
     \left \Vert   \varphi_{m,h}
     -  \mathfrak{g}_{m,h}\right\Vert^2_{L^2(\cQ)}
     \leq \frac{\tilde{\epsilon}}{6 \pi c  2^{|h|}},
     \quad  -L \leq h \leq L, \; m =1,2 \ldots, M.
\end{align}
Note that the existence of such networks is guaranteed
by Assumption \ref{as:sg}. For each $m=1,2,\ldots,M$
and  any positive integer $L$, the finite sequence
$\{\mathfrak{g}_{m,h}\}_{-L \leq h \leq L}$ is extended
to  an  infinite sequence $\{\mathfrak{g}_{m,h}\}_{h \in \mathbb{Z}}$
in  $\mathcal{C}$ by setting the remaining  elements
to zero.  The  Fourier transform of the series
$\{\mathfrak{g}_{m,h}\}_{h \in \mathbb{Z}}$
is denoted by $\mathfrak{g}^{\ddagger}_{m}( \theta) $.  We use notation $\cdot^{\ddagger}$ to emphasize this is indeed the Fourier transform of a finite series.
In particular,  $\mathfrak{g}^{\ddagger}_{m}( \theta) $ and
$\mathfrak{g}_{m,h}$  are connected through Definition \ref{def:FT}.
Observe that, for each $m=1, 2, \ldots, M$ and positive integer $L$, we have
\begin{align}
 \label{eq:phi-f}
   & \frac{1}{4}  \int_{-\pi}^{\pi}\left \Vert \varphi^{\dagger}_m(\theta)
   - \mathfrak{g}^{\ddagger}_{m}( \theta) \right\Vert^2_{L^2(\cQ)} d \theta\\ \label{eq:phi}
   \leq & \int_{-\pi}^{\pi} \left \Vert \varphi^{\dagger}_m(\theta)  -  \sum_{h= -L}^L \exp(\mi h \theta) \varphi_{m,h}\right\Vert^2_{L^2(\cQ)} d \theta\\ \label{eq:phi:f}
   &+ \int_{-\pi}^{\pi}
   \left \Vert \sum_{h= -L}^L \exp (\mi h \theta) \varphi_{m,h} - \sum_{h= -L}^L  \exp(\mi h \theta) \mathfrak{g}_{m,h}\right\Vert^2_{L^2(\cQ)} d \theta\\ \label{eq:f}
  & + \int_{-\pi}^{\pi}
    \left \Vert \sum_{h= -L}^L  \exp (\mi h \theta) \mathfrak{g}_{m,h}
    - \mathfrak{g}^{\ddagger}_{m}( \theta) \right\Vert^2_{L^2(\cQ)} d \theta.
 \end{align}
We now prove that there exists sufficiently large $L$
such the above summands can be bounded above by
arbitrarily small $\tilde{\epsilon} >0$.
First observe that, relation \eqref{eq:FT} guarantees that  there exists a sufficiently large $L= L(M)$   such that  \eqref{eq:phi} is bounded by $\tilde{\epsilon}$, for $m=1,2\ldots,M$.

For this $L$, inequality \eqref{eq:nn:prf} implies that  \eqref{eq:phi:f} is upper bounded by
\begin{align*}
\int_{-\pi}^{\pi}  \left \Vert \sum_{h= -L}^{L} \exp(\mi h \theta) \left( \varphi_{m,h} -  \mathfrak{g}_{m,h} \right)\right\Vert^2_{L^2(\cQ)} d \theta
   \\
   \leq & c \int_{-\pi}^{\pi}
\sum_{h= -L}^{L}   \left \Vert  \varphi_{m,h} -  \mathfrak{g}_{m,h} \right\Vert^2_{L^2(\cQ)} d \theta  \\
 \leq & c \int_{-\pi}^{\pi}
\sum_{h= -L}^{L}   \frac{\tilde{\epsilon}}{6 \pi c  2^{|h|}} d \theta   \\
   \leq &    2   \pi c  \sum_{h= -\infty}^{\infty} \frac{\tilde{\epsilon}}{6 \pi c  2^{|h|}} =  \frac{ 2   \pi c \tilde{\epsilon}}{6 \pi c } \times 3 = \tilde{\epsilon}.
\end{align*}
By construction, \eqref{eq:f} equals zero.

Summarizing the argument above,
relation \eqref{eq:FT} implies that  there is  a
sufficiently large positive integer $L$ for which \eqref{eq:phi}
is bounded by arbitrarily small $\tilde{\epsilon}$.
For this finite $L$, there exist finite sequences of the neural
networks $\{\mathfrak{g}_{m,h}\}_{-L \leq h \leq L}$
satisfying \eqref{eq:nn:prf}. This implies  \eqref{eq:phi:f}
is bounded above by $\tilde{\epsilon}$. The finite sequences
$\{\mathfrak{g}_{m,h}\}_{-L \leq h \leq L}$ are  extended to
infinite sequences $\{\mathfrak{g}_{m,h}\}_{h \in \mathbb{Z}}$
such that \eqref{eq:f} equals zero.
Consequently, there exist sufficiently large $L$
and $\mathfrak{g}^{\ddagger}_{m}( \theta)  \in \mathcal{D}$,
for $m=1,\ldots M$, such that \eqref{eq:phi-f} satisfies
\begin{align}\label{e:p-f:dg}
     \int_{-\pi}^{\pi}\left \Vert \varphi^{\dagger}_m(\theta)  - \mathfrak{g}^{\ddagger}_{m}( \theta) \right\Vert^2_{L^2(\cQ)} d \theta \leq 8 \tilde{\epsilon}  , \quad m=1,2,\ldots,M .
\end{align}
Now observe that
\begin{align}
\nonumber
    &  \left \Vert \sum_{m =1}^{M} \lambda_m(\theta) \varphi^{\dagger}_m(\theta)\otimes \varphi^{\dagger}_m(\theta) - \sum_{m =1}^{M} \lambda_m(\theta)  \mathfrak{g}^{\ddagger}_{m}( \theta) \otimes \mathfrak{g}^{\ddagger}_{m}( \theta)
   \right \Vert_{\mathcal{S}} \\ \label{eq:||L^2}
   \leq & \sum_{m =1}^{M}  2 \lambda_m(\theta)
   \left \Vert \varphi^{\dagger}_m(\theta)  - \mathfrak{g}^{\ddagger}_{m}( \theta) \right\Vert_{L^2(\cQ)}
   +   \sum_{m =1}^{M}  \lambda_m(\theta) \left \Vert \varphi^{\dagger}_m(\theta) - \mathfrak{g}^{\ddagger}_{m}( \theta) \right\Vert^2_{L^2(\cQ)}\\ \label{eq:s1+s2}
   \leq & 2 \Lambda^{\ast} \sum_{m =1}^{M}
   \left \Vert \varphi^{\dagger}_m(\theta) - \mathfrak{g}^{\ddagger}_{m}( \theta) \right\Vert_{L^2(\cQ)} + \Lambda^{\ast} \sum_{m =1}^{M}   \left \Vert \varphi^{\dagger}_m(\theta) - \mathfrak{g}^{\ddagger}_{m}( \theta) \right\Vert^2_{L^2(\cQ)},
\end{align}
 where   $ \Lambda^{\ast} =   \underset{m,\theta}{\sup}\; \lambda_m(\theta) < \infty$ is defined in Lemma \ref{lem:sum:lbd} and inequality \eqref{eq:||L^2} is a consequence of $\Vert f \otimes f - g \otimes g \Vert_{L^2 (\cQ \times \cQ)} \leq 2 \Vert f \Vert \Vert f-g \Vert + \Vert f-g \Vert^2$.
 Then, for $\tilde{\epsilon}$ sufficiently small, \eqref{eq:s1+s2} and \eqref{e:p-f:dg} imply
 \begin{align}
\nonumber
    & \int_{-\pi}^{\pi} \left \Vert \sum_{m =1}^{M} \lambda_m(\theta) \varphi^{\dagger}_m(\theta)\otimes \varphi^{\dagger}_m(\theta) - \sum_{m =1}^{M} \lambda_m(\theta)  \mathfrak{g}^{\ddagger}_{m}( \theta) \otimes \mathfrak{g}^{\ddagger}_{m}( \theta)
   \right \Vert_{\mathcal{S}} d \theta \\ \nonumber
   \leq & 2 \Lambda^{\ast}\sum_{m =1}^{M} \left[  \int_{-\pi}^{\pi}
   \left \Vert \varphi^{\dagger}_m(\theta) - \mathfrak{g}^{\ddagger}_{m}( \theta) \right\Vert^{2}_{L^2(\cQ)}  d \theta \right]^{1/2}\\ \nonumber
   &+ \Lambda^{\ast} \sum_{m =1}^{M}   \int_{-\pi}^{\pi}
   \left \Vert \varphi^{\dagger}_m(\theta)  - \mathfrak{g}^{\ddagger}_{m}( \theta) \right\Vert^{2}_{L^2(\cQ)}  d \theta \\ \label{e:fM-f}
   \leq &  3 \Lambda^{\ast}  M (8\tilde{\epsilon})^{1/2}.
\end{align}
Setting $\mathfrak{f}(\theta)  = \sum_{m =1}^{M} \lambda_m(\theta)  \mathfrak{g}^{\ddagger}_{m}( \theta) \otimes \mathfrak{g}^{\ddagger}_{m}( \theta) \in \mathcal{E}$ and
choosing  $\tilde{\epsilon}$ sufficiently small, as a function of $\epsilon$, \eqref{e:fM-f} entails
\begin{align}\label{eq:s2}
    \int_{-\pi}^{\pi} \left \Vert f_{M}^{X}(\theta)  - \mathfrak{f}(\theta)
   \right \Vert_{\mathcal{S}} d \theta \leq \epsilon/2.
\end{align}
Combining inequalities \eqref{eq:fX-f}, \eqref{eq:s1} and \eqref{eq:s2},
we obtain the desired universal approximation.
\hfill \QED

\bigskip
\noindent
{\sc Proof of Theorem \ref{thm:E:fX}.}
\noindent \textbf{Step 1:}
In this step, we prove that  the spectral density kernel of the stationary process
$ \{\widetilde{\mathfrak{X}}_t\}$ defined in the statement of Theorem  \ref{thm:E:fX}  has the representation  \eqref{eq:F:fX}.
To do so, rewrite $\widetilde{\mathfrak{X}}_t $ in the form
\begin{align*}
    \widetilde{\mathfrak{X}}_t
    = &
   \sum_{h = -L}^L  \left( \mathfrak{g}_{1, h} , \ldots, \mathfrak{g}_{M, h} \right) \left(\xi_{1,t+h} , \ldots, \xi_{M,t+h}\right)^{\top}\\
    =: &  \sum_{h = -L}^L \mathfrak{g}_{h} \xi^{\top}_{t+h}.
\end{align*}
The spectral density operator of the stationary random process $\{\widetilde{\mathfrak{X}}_t\}$ has the form
\begin{align}
\nonumber
    F^{\widetilde{\mathfrak{X}}}(\theta) = & \frac{1}{2 \pi } \sum_{h \in \mathbb{Z}} C^{\widetilde{\mathfrak{X}}}_h \exp (-\mi h \theta)\\ \nonumber
    =&  \frac{1}{2 \pi } \sum_{h \in \mathbb{Z}}
    \mathrm{Cov} (\widetilde{\mathfrak{X}}_h, \widetilde{\mathfrak{X}}_0)
    \exp (-\mi h \theta)\\ \nonumber
    =&  \frac{1}{2 \pi } \sum_{h \in \mathbb{Z}}
    \mathrm{Cov} (
    \sum_{s =-L}^L  \mathfrak{g}_{s} \xi^{\top}_{h+s}
    ,
   \sum_{s' =-L}^L  \mathfrak{g}_{s'} \xi^{\top}_{s'}
    )
    \exp (-\mi h \theta).
    \end{align}
 This implies
   \begin{align}
    \nonumber
  F^{\widetilde{\mathfrak{X}}}(\theta) =&  \frac{1}{2 \pi } \sum_{h \in \mathbb{Z}} \sum_{s =-L}^L  \sum_{s' =-L}^L
    \mathrm{Cov} (
     \mathfrak{g}_{s} \xi^{\top}_{h+s}
    ,
   \mathfrak{g}_{s'} \xi^{\top}_{s'}
    )
    \exp (-\mi h \theta)\\    \nonumber
    =&  \frac{1}{2 \pi } \sum_{h \in \mathbb{Z}} \sum_{s =-L}^L  \sum_{s' =-L}^L
 \mathfrak{g}_{s}   \mathrm{Cov} (
      \xi^{\top}_{h+s}
    ,
    \xi^{\top}_{s'}
    ) \overline{\mathfrak{g}}^{\top}_{s'}
    \exp (-\mi h \theta -\mi s \theta +\mi s' \theta+\mi s \theta -\mi s' \theta ) .
\end{align}
By the summability conditions $\sum_{h \in \mathbb{Z}} \Vert C_h^{\xi}\Vert_{\mathcal{S}} < \infty$,
we have
    \begin{align*}
   F^{\widetilde{\mathfrak{X}}}(\theta) =& \frac{1}{2 \pi }  \sum_{s \in \mathbb{Z}} \sum_{s' \in \mathbb{Z}}   \sum_{h \in \mathbb{Z}}
 \mathfrak{g}_{s}   \mathrm{Cov} (
      \xi^{\top}_{h+s}
    ,
    \xi^{\top}_{s'}
    ) \overline{\mathfrak{g}}^{\top}_{s'}
    \exp (-\mi h \theta -\mi s \theta +\mi s' \theta+\mi s \theta -\mi s' \theta )\\
 =& \frac{1}{2 \pi }  \sum_{s =-L}^L \sum_{s' =-L}^L
 \mathfrak{g}_{s}
  F^{\xi}(\theta)\overline{\mathfrak{g}}^{\top}_{s'}
    \exp ( \mi s \theta -\mi s' \theta ).
\end{align*}
This gives the desired form \eqref{eq:F:fX}.

\bigskip
\noindent
\textbf{Step 2:}
In this step, we prove that the class  $\mathcal{E}$ defined in  \eqref{eq:cF} can be written in the form \eqref{e:alt-f}.
Consider a generic element $\mathfrak{f}$ in the class $\mathcal{E}$ defined in  \eqref{eq:cF} given by
\begin{align*}
    &\mathfrak{f} (\theta ) =  \sum_{m=1}^M
     \eta_{m} (\theta)
     \mathfrak{g}^{\ddagger}_{m}( \theta) \otimes \mathfrak{g}^{\ddagger}_{m} (\theta), \quad \theta \in [-\pi , \pi],
\end{align*}
for some   $M \in \mathbb{N}$,
$\eta_{\cdot} (\cdot) \in \mathcal{A}_{\{1,\ldots,M\}} \subset \mathcal{A}$,
$\mathfrak{g}^{\ddagger}_{m}  \in  \mathcal{D}$, $m=1, \ldots M$.
According to Step 1, it is enough to prove the existence of an $M$-dimensional random process  $\{\xi_t = \left( \xi_{1,t}, \ldots , \xi_{M,t}\right)\}$ with the spectral density operator $\mathrm{diag}\left( \eta_1(\theta), \ldots, \eta_M(\theta)\right) $. Consider the Gaussian $M$-dimensional random process  $\{\xi_t = \left( \xi_{1,t}, \ldots , \xi_{M,t}\right)\}$ with independent component and the following covariance structure for its components:
\begin{align*}
    c^m_h = \int_{- \pi}^{\pi}  \eta_m(\theta) \exp (\mi h \theta) d \theta, \quad m=1, \ldots M.
\end{align*}
Since $\eta_{\cdot} (\cdot) \in \mathcal{A}$,
the above covariances are well-defined. Since they form a positive-definite family,  the existence of the Gaussian process $\xi_t$ follows. See e.g. Chapter 1 of \citetext{brockwell:davis:1991}.
This completes the proof.
\hfill \QED

\begin{remark}
Step 1 in the proof of Theorem \ref{thm:E:fX} could  also be derived from  Theorem 2.5.5 in \citetext{tavakoli:2014}.
   Theorem 2.5.5 in \citetext{tavakoli:2014} imposes two assumptions: their Condition 2.4.1($p$) for some $p \in [1, \infty)$ and the limiting relation (2.5.12). In our case we have the summability
   assumption $\sum_{h \in \mathbb{Z}} \Vert C_h^{\xi} \Vert_{\mathcal{S}} < \infty$. This summability condition implies $\sum_{h \in \mathbb{Z}} \Vert C_h^{\xi} \Vert_{\mathcal{N}} < \infty$, where $\Vert \cdot \Vert_{\mathcal{N}}$ denotes the nuclear norm.
   This follows because all norms defined in finite-dimensional topological vector spaces are equivalent. Consequently, $\sum_{h \in \mathbb{Z}} \Vert C_h^{\xi} \Vert_{\mathcal{N}} < \infty$ implies  Condition 2.3.3 and Condition 2.3.4 in \citetext{tavakoli:2014}. Following their Remark 2.4.2, we conclude Condition 2.4.1 for $p = \infty$.
Additionally, in our Theorem \ref{thm:E:fX},
   we work with finite sequences $\{\mathfrak{g}_h\}_{-L \leq h \leq L}$ and in particular  the summability condition $\sum_{h \in \mathbb{Z}}
   \Vert \mathfrak{g}_{m,h}\Vert_{L^2(\cQ)} < \infty$ holds.
   According to Remark 2.5.6 in \citetext{tavakoli:2014},
   this implies  their formula (2.5.12). In summary, the assumptions of
   Theorem 2.5.5 in \citetext{tavakoli:2014} hold in our case.
   Therefore,
   the form of the spectral density operator of
   $\{\widetilde{\mathfrak{X}}_t\}$ is a  consequence of Theorem 2.5.5 in
   \citetext{tavakoli:2014}.
\end{remark}

Before proceeding with the proof of Theorem \ref{thm:wq}, we review and modify for our purposes the
required background on functional filtered processes. For a more comprehensive discussion, we refer to Sections
A.3 and A.4 of \citetext{hormann:dynamic:2015}. Recall the discussion at the beginning of Section \ref{p:spec}.
The following lemma follows from  calculations in Subsection
A.4.1 of  \citetext{hormann:dynamic:2015}.

\begin{lemma}\label{lem:X-gamY}
Suppose Assumption \ref{as:sum:Ch} holds and
consider an array of functions $\ga_{m,l} \in L^2(\cQ)$ (a sequence of
linear filters) such  that
\[
\forall \ m \geq 1, \ \ \sum_{l\in \mbZ} \|\ga_{m,l} \| < \infty.
\]
For the $\varphi_{m,l}$ in  \eqref{eq:varphi} and the
$Y_{m,t}$ in \eqref{eq:X:inv}, set
\[
\gamma^{(Y)}_{M,t}
 = \sum_{m=1}^{M} \sum_{l \in \mathbb{Z}} Y_{m, t+l} \gamma_{m,l}.
\]
Then the series $\gamma^{(Y)}_{M,t}$ is well-defined in  $L^2(\cQ)$
and
 \begin{align*}
     \mathbb{E} \left\Vert X_t - \gamma^{(Y)}_{M,t} \right\Vert^2
     = \int_{-\pi}^{\pi}  \left\Vert
     \sqrt{F^{X}(\theta)} - \Gamma_M(\theta)\sqrt{F^{X}(\theta)}
     \right\Vert_{\mathcal{S}}^2   d\theta,
 \end{align*}
 where  $\sqrt{F^{X}(\theta)}$ denotes the square
 root of $F^{X}(\theta)$ and
\[
\Gamma_M(\theta)
= \sum_{m=1}^{M} \left[\sum_l \gamma_{m,l}\exp{(\mi l \theta)}  \right]
\otimes \varphi^{\dagger}_m(\theta).
\]
\end{lemma}

We now turn to a mean-square network  approximation of the  $X_t$.

\begin{proposition}\label{prop:long2}
Suppose  Assumptions \ref{as:sum:Ch} and \ref{as:sg} hold.
Then, there are networks   $\{\widetilde{\mathfrak{X}}_t\}$
as in Assumption \ref{a:sta}, indexed by $M,L$, such that for each $t\in \mbZ$,
\begin{align*}
\mathbb{E} \Vert X_t - \widetilde{\mathfrak{X}}_t \Vert^2 \rightarrow 0,
\; \; \text{as }  M,L \to  \infty.
\end{align*}
\end{proposition}
{\sc Proof.}
Recall the spectral density decomposition \refeq{FXdec}.
Equations \eqref{eq:ph^2} and \eqref{eq:varphi}
imply that, for each pair $(m,l)$,  $\varphi_{m,l} \in L^2(\cQ)$.
Assumption \ref{as:sg} then implies that
there are approximating neural networks
$\mathfrak{g}_{m,l} \in \mathcal{C}^{\text{\tiny nn}}$
arbitrarily  close to $\varphi_{m,l} $ in  the $L^2(\cQ)$ distance.
For finite positive integers $L$ and $M$, to be defined later,
choose the networks $\mathfrak{g}_{m,l}$ such that \eqref{eq:nn:prf}
holds. And again,  let
$\mathfrak{g}_m^{\ddagger} (\theta)$ be the
Fourier transform of the finite sequence
$\{\mathfrak{g}_{m,l}\}_{-L \leq l \leq L}$.
For $M,L \geq 1$, define
\[
\widetilde{\mathfrak{X}}_t =
\widetilde{\mathfrak{X}}_t^{M,L}
=   \sum_{m=1}^{M} \sum_{l =-L}^L Y_{m, t+l} \mathfrak{g}_{m,l},
\]
where the $Y_{m,t}$ are defined in \eqref{eq:X:inv}.
It is enough to show that for an arbitrary small $\epsilon >0$
there exist   sufficiently large $M$ and $L$ such that
\begin{align}\label{eq:ms:X-tX}
    \mathbb{E} \Vert X_t - \widetilde{\mathfrak{X}}_t \Vert^2
   < \epsilon.
\end{align}
To obtain \eqref{eq:ms:X-tX}, we will apply Lemma \ref{lem:X-gamY}.
Observe first that
\begin{align*}
     \Gamma_M(\theta) = &\sum_{m=1}^{M} \left[\sum_{l=-L}^{L} \mathfrak{g}_{m,l}\exp{(\mi l \theta)}  \right] \otimes \varphi^{\dagger}_m(\theta) \\
    =& \sum_{m=1}^{M} \mathfrak{g}^{\ddagger}_{m}( \theta)\otimes \varphi^{\dagger}_m(\theta) .
\end{align*}
Recall that    $\mathfrak{g}_m^{\ddagger} (\theta)$
and $\mathfrak{g}_{m,l}$ are connected through Definition \ref{def:FT}.
Now, observe that
\begin{align*}
\Gamma_M(\theta)\sqrt{F^{X}(\theta)} = &  \left[\sum_{m=1}^{M} \mathfrak{g}^{\ddagger}_{m}( \theta)\otimes  \varphi^{\dagger}_m(\theta) \right]\left[\sum_{m \geq 1} \sqrt{\lambda_m(\theta) }\varphi^{\dagger}_m(\theta)\otimes \varphi^{\dagger}_m(\theta)\right]\\
    =& \sum_{m=1}^{M} \sum_{m' \geq 1} \sqrt{\lambda_{m'}(\theta) }\langle \varphi^{\dagger}_{m'}, \varphi^{\dagger}_{m} \rangle \mathfrak{g}^{\ddagger}_{m}( \theta)\otimes  \varphi^{\dagger}_{m'}(\theta) \\
    =& \sum_{m=1}^{M} \sqrt{\lambda_m(\theta)} \mathfrak{g}^{\ddagger}_{m}( \theta)\otimes  \varphi^{\dagger}_{m}(\theta).
\end{align*}
Therefore,
\begin{align*}
     \sqrt{F^{X}(\theta)} -  \Gamma_M
     (\theta)\sqrt{F^{X}(\theta)} =& \sum_{m \geq 1} \sqrt{\lambda_m(\theta) }\varphi^{\dagger}_m(\theta)\otimes\varphi^{\dagger}_m(\theta) - \sum_{m=1}^{M} \sqrt{\lambda_m(\theta)} \mathfrak{g}^{\ddagger}_{m}( \theta)\otimes  \varphi^{\dagger}_{m}(\theta) \\
     =& \sum_{m \geq 1} \sqrt{\lambda_m(\theta) }\varphi^{\dagger}_m(\theta)\otimes\varphi^{\dagger}_m(\theta) - \sum_{m=1}^{M} \sqrt{\lambda_m(\theta)} \varphi^{\dagger}_{m}(\theta)\otimes  \varphi^{\dagger}_{m}(\theta)\\
     &+ \sum_{m=1}^{M} \sqrt{\lambda_m(\theta)} \varphi^{\dagger}_{m}( \theta)\otimes  \varphi^{\dagger}_{m}(\theta) - \sum_{m=1}^{M} \sqrt{\lambda_m(\theta)} \mathfrak{g}^{\ddagger}_{m}( \theta)\otimes  \varphi^{\dagger}_{m}(\theta).
\end{align*}
Hence,
\begin{align*}
  \frac{1}{2}\left\Vert
     \sqrt{F^{X}(\theta)} -  \Gamma_M
     (\theta)\sqrt{F^{X}(\theta)}
     \right\Vert_{\mathcal{S}}^2 \leq &    \left\Vert
    \sum_{m > M} \sqrt{\lambda_m(\theta) }\varphi^{\dagger}_m(\theta)\otimes\varphi^{\dagger}_m(\theta)
     \right\Vert_{\mathcal{S}}^2\\
     & +   \left\Vert
     \sum_{m=1}^{M} \sqrt{\lambda_m(\theta)} \left[\varphi^{\dagger}_{m}( \theta) - \mathfrak{g}^{\ddagger}_{m}( \theta)\right]\otimes  \varphi^{\dagger}_{m}(\theta)
     \right\Vert_{\mathcal{S}}^2\\
     =& \sum_{m > M} \lambda_m(\theta)
     +  \sum_{m=1}^{M} \lambda_m(\theta) \Vert \varphi^{\dagger}_{m}( \theta) - \mathfrak{g}^{\ddagger}_{m}( \theta) \Vert^2 .
\end{align*}
This implies
\begin{align*}
    \frac{1}{2}\mathbb{E} \Vert X_t - \widetilde{\mathfrak{X}}_t \Vert^2 \leq &
    \int_{-\pi}^{\pi}
\sum_{m > M} \lambda_m(\theta)
     d\theta
     +
\Lambda^{\ast}
 \sum_{m =1}^{M}   \int_{-\pi}^{\pi}
   \left \Vert \varphi^{\dagger}_m(\theta)  - \mathfrak{g}^{\ddagger}_{m}( \theta) \right\Vert^{2}
         d\theta
           \\
           = & \sum_{m > M} \Lambda_m + \Lambda^{\ast}
 \sum_{m =1}^{M}   \int_{-\pi}^{\pi}
   \left \Vert \varphi^{\dagger}_m(\theta)  - \mathfrak{g}^{\ddagger}_{m}( \theta) \right\Vert^{2}
         d\theta
           \\
    =:&  S_1 + S_2 .
\end{align*}
where   $ \Lambda^{\ast} =   \underset{m,\theta}{\sup}\; \lambda_m(\theta) < \infty$ is defined in Lemma \ref{lem:sum:lbd}.
Lemma \ref{lem:Lbd} guarantees that there is a sufficiently large
 $M$ such that $S_1 < \epsilon/4$.
For this $M$, an argument similar to that leading to \eqref{e:p-f:dg}, implies that for a sufficiently large
 $L$, $S_2 < \epsilon/4$.
This completes the proof.
\hfill \QED

\begin{lemma}\label{lem:c_h}
Consider the setting of Proposition \ref{prop:long2}. Recall that  the lag $h$ autocovariance operators of the
stationary processes $\{\widetilde{\mathfrak{X}}_t^{M,L}\}$
and $\{X_t\}$  are denoted by  $C_h^{\widetilde{\mathfrak{X}}}$
and $C_h^X$, respectively. Then,
\begin{equation} \label{e:MLh}
\lim_{M, L\to \infty}
\sup_{h \in \mbZ} \left \Vert C_h^{\widetilde{\mathfrak{X}}}
  - C_h^X\right \Vert_{\mathcal{S}}^2 = 0.
\end{equation}
\end{lemma}

\bigskip
\noindent
{\sc Proof of Lemma \ref{lem:c_h}}
Observe that
\begin{align*}
     C_h^{\widetilde{\mathfrak{X}}} - C_h^X & =    \mathbb{E} [\widetilde{\mathfrak{X}}_h \otimes \widetilde{\mathfrak{X}}_0] - \mathbb{E} [X_h \otimes X_0]
    \\
    & =  \mathbb{E} [(\widetilde{\mathfrak{X}}_h -X_h) \otimes (\widetilde{\mathfrak{X}}_0-X_0)]
    + \mathbb{E} [(\widetilde{\mathfrak{X}}_h -X_h) \otimes X_0]
    + \mathbb{E} [X_h \otimes (\widetilde{\mathfrak{X}}_0-X_0)].
\end{align*}
This implies
\begin{align*}
    \left \Vert C_h^{\widetilde{\mathfrak{X}}} - C_h^X\right \Vert_{\mathcal{S}}
    \leq &
  \mathbb{E}  \left \Vert  \widetilde{\mathfrak{X}}_h -X_h \right \Vert \left \Vert  \widetilde{\mathfrak{X}}_0-X_0
    \right \Vert
    + \mathbb{E} \left \Vert \widetilde{\mathfrak{X}}_h -X_h \right \Vert \left \Vert  X_0
    \right \Vert
    + \mathbb{E} \left \Vert  X_h \right \Vert \left \Vert \widetilde{\mathfrak{X}}_0-X_0
    \right \Vert.
\end{align*}
Consequently,
\begin{align*}
\frac{1}{4}  &
  \left \Vert C_h^{\widetilde{\mathfrak{X}}}
  - C_h^X\right \Vert_{\mathcal{S}}^2\\
 & \leq
 \mathbb{E}   \left \Vert  \widetilde{\mathfrak{X}}_h -X_h \right \Vert^2 \mathbb{E} \left \Vert  \widetilde{\mathfrak{X}}_0-X_0
    \right \Vert^2
    +
    \mathbb{E} \left \Vert  \widetilde{\mathfrak{X}}_h -X_h \right \Vert^2\mathbb{E} \left \Vert  X_0
    \right \Vert^2
    +
    \mathbb{E}\left \Vert  X_h  \right \Vert^2 \mathbb{E} \left \Vert \widetilde{\mathfrak{X}}_0-X_0
    \right \Vert^2,
\end{align*}
and so, by Proposition \ref{prop:long2},
\begin{equation*} 
\lim_{M, L\to \infty}
\sup_{h \in \mbZ} \left \Vert C_h^{\widetilde{\mathfrak{X}}}
  - C_h^X\right \Vert_{\mathcal{S}}^2 = 0
\end{equation*}
as desired.
\hfill \QED

\bigskip
\noindent
 {\sc Proof of Theorem \ref{thm:wq}.} Using the triangle inequality, for any $q \geq 1$,  we have
    \begin{align*}
   &      \int_{-\pi}^{\pi} \Vert f^{X}(\theta)  -
         \sum_{|h|\leq q}     \omega\lp\frac{h}{q}\rp C^{{\mathfrak{X}}}_h \exp (-\mi h \theta) \Vert_{\mathcal{S}} d \theta \\
         \leq & \int_{-\pi}^{\pi} \Vert f^{X}(\theta)  -
         \sum_{|h|\leq q}     C^{X}_h \exp (-\mi h \theta) \Vert_{\mathcal{S}} d \theta \\
         & + \int_{-\pi}^{\pi} \Vert  \sum_{|h|\leq q}   C^{X}_h \exp (-\mi h \theta)  -
         \sum_{|h|\leq q}     \omega\lp\frac{h}{q}\rp C^{X}_h \exp (-\mi h \theta) \Vert_{\mathcal{S}} d \theta \\
         & + \int_{-\pi}^{\pi} \Vert  \sum_{|h|\leq q}     \omega\lp\frac{h}{q}\rp C^{X}_h \exp (-\mi h \theta)   -
         \sum_{|h|\leq q}     \omega\lp\frac{h}{q}\rp C^{{\mathfrak{X}}}_h \exp (-\mi h \theta) \Vert_{\mathcal{S}} d \theta\\
         =: & A_1 + A_2 + A_3.
    \end{align*}
It is now enough to show that each of the summands can be bounded by $\epsilon/3$. For $A_1$, observe that by H{\"o}lder's inequality and the definition of the Hilbert–Schmidt norm, we have
\begin{align*}
  A_1^2 \leq &  \int_{-\pi}^{\pi} \Vert f^{X}(\theta)  -
         \sum_{|h|\leq q}     C^{X}_h \exp (-\mi h \theta) \Vert^2_{\mathcal{S}} d \theta \\
     = & \int_{-\pi}^{\pi} \Vert
         \sum_{|h| > q}     C^{X}_h \exp (-\mi h \theta) \Vert^2_{\mathcal{S}} d \theta \\
         = & \int_{-\pi}^{\pi}   \iint_{\cQ \times \cQ}
       \left\vert  \sum_{|h| > q}     c^{X}_h (u,v) \exp (-\mi h \theta) \right\vert^2 du dv d \theta
       \end{align*}
An application of Parseval’s equality implies
      \begin{align*}
       A_1^2 \leq  & 2 \pi  \sum_{|h| > q} \iint_{\cQ \times \cQ}
       \left\vert      c^{X}_h (u,v)  \right\vert^2 du dv  \\
       = & 2 \pi  \sum_{|h| > q} \Vert C_h^X \Vert^2_{\mathcal{S}}.
\end{align*}
According to Assumption \ref{as:sum:Ch}, there exists a sufficiently large $q$ such that the sum above is bounded by $\epsilon^2/9$, i.e $A_1 < \epsilon/3$.

The argument for $A_2$ depends on the kernel being used,
but it is clear that it will work for any kernel used in practice.
Observe that
\begin{align*}
    A_2 = \int_{-\pi}^{\pi} \left \Vert \sum_{|h|\leq q}     \left (1-  \omega(h/q)\right ) C^X_h  \exp (-\mi h \theta) \right \Vert_{\mathcal{S}} d \theta
    \leq 2 \pi \sum_{|h|\leq q} \left \vert 1-  \omega(h/q) \right \vert
    \left \Vert C^X_h \right \Vert_{\mathcal{S}} .
\end{align*}
Hence, using the Truncated kernel, $A_2 =0 $. Using the Bartlett kernel
\begin{align*}
  \frac{1}{2 \pi }  A_2 \leq  &   \sum_{|h|\leq q} \left \vert 1-  \left(1- \frac{\vert h \vert}{q} )\right) \right \vert
    \left \Vert C^X_h \right \Vert_{\mathcal{S}}
    =
     \sum_{|h|\leq q} \frac{\vert h \vert }{q}
    \left \Vert C^X_h \right \Vert_{\mathcal{S}} \\
    \leq & \sum_{|h|\leq q} \frac{\vert h \vert^{\alpha} }{q^{\alpha} }
    \left \Vert C^X_h \right \Vert_{\mathcal{S}}
    \leq \frac{1}{q^{\alpha}}  \sum_{h \in \mathbb{Z}} \vert h \vert^{\alpha}
    \left \Vert C^X_h \right \Vert_{\mathcal{S}},
\end{align*}
where, by Assumption \ref{a:3k}, $\sum_{h \in \mathbb{Z}} \vert h \vert^{\alpha}
\left \Vert C^X_h \right \Vert_{\mathcal{S}} < \infty$.
Assumption \ref{a:3k} also implies
$q^{\alpha}$ diverges to infinity and hence for sufficiently large $q$,
 the term $A_2$ is bounded by $\epsilon/3$. Using the Parzen kernel,  we have
    \begin{align*}
\frac{1}{2 \pi }    A_2
    \leq &  \sum_{|h|=0}^{\frac{q}{2}-1} \left \vert
   1- 1 + 6 \frac{\vert h \vert^2}{q^2} - 6 \frac{\vert h \vert^3}{q^3}
    \right \vert
    \left \Vert C^X_h \right \Vert_{\mathcal{S}} + \sum_{|h|=\frac{q}{2}}^{q}
    \left \vert
   1- 2 \left( 1- \frac{\vert h \vert}{q}\right)^3
    \right \vert
    \left \Vert C^X_h \right \Vert_{\mathcal{S}} \\
    \leq &
   6 \sum_{|h|=0}^{\frac{q}{2}-1} \left \vert
  \frac{\vert h \vert^2 q ^3 - \vert h \vert^3 q ^2 }{q^5}
    \right \vert
    \left \Vert C^X_h \right \Vert_{\mathcal{S}}  + \sum_{|h|=\frac{q}{2}}^{q}
   \left (
  1 + 2
    \right )
    \left \Vert C^X_h \right \Vert_{\mathcal{S}}\\
    \leq &
    \frac{6}{q^{\alpha}} \sum_{h \in \mathbb{Z}}
    \vert h \vert^{\alpha}
    \left \Vert C^X_h \right \Vert_{\mathcal{S}}  + 3\sum_{|h|=\frac{q}{2}}^{\infty}
    \left \Vert C^X_h \right \Vert_{\mathcal{S}}.
\end{align*}
The first term can be made arbitrarily small by
Assumption \ref{a:3k} and the second by Assumption \ref{as:sum:Ch}.

We now turn to the last term $A_3$, for which we only need the fact that $\og$ is bounded. This follows from the specific choice of functions in Assumption \ref{a:3k}, or more generally, for example, for any continuous and compactly supported kernel.
Choose the bandwidth $q$ sufficiently large such that $A_1$ and $A_2$
are bounded by $\epsilon/3$.
Since  the weight function $\omega(\cdot)$ is bounded
by some finite constant $c$,
\begin{align*}
    A_3 \leq & \int_{-\pi}^{\pi} \sum_{|h|\leq q}    \left\vert \exp (-\mi h \theta) \omega\lp\frac{h}{q}\rp \right\vert  \Vert     C^{X}_h   -
              C^{{\mathfrak{X}}}_h  \Vert_{\mathcal{S}} d \theta \\
              \leq & c \int_{-\pi}^{\pi} \sum_{|h|\leq q}    \Vert     C^{X}_h   -
              C^{{\mathfrak{X}}}_h  \Vert_{\mathcal{S}} d \theta
         \\ \leq &  2 \pi c \sum_{|h|\leq q}
    \left \Vert
 C^X_h   - C^{\widetilde{\mathfrak{X}}}_h
    \right \Vert_{\mathcal{S}}.
\end{align*}
Relation \refeq{MLh} implies that
for sufficiently large $M$  and $L$,  the term $A_3$ is bounded by $\epsilon/3$.
This completes the proof.
\hfill \QED

\section{Additional simulation results}\label{sec:additional_simulations}
In this section, we report the simulation results in the 1D and 3D setups mentioned in Section~\ref{s:sim}. In Tables~\ref{tab:1D_N_K} and \ref{tab:1D_gamma}, we report the estimation errors in 1D. The estimation errors in 3D are shown in Table~\ref{tab:3D_N_K}. We also report the computing times and maximum memory requirements by the two estimators in Table~\ref{tab:1D_time_memory} for 1D and Table~\ref{tab:3D_time_memory} for 3D. For the spectral-NN estimator, we also report the computing times with GPU computing. The time and memory complexities are obtained from runs on a computer with 64 GiB RAM, AMD Ryzen 9 5900X (3.7 GHz) CPU, NVIDIA GeForce RTX 3090 GPU, and Ubuntu 24.04.2 LTS (64-bit) OS.

Our observations in the 1D and 3D setups remain similar to those made in the 2D setup. The estimation error for both the estimators decrease with increasing sample sizes as well as increasing resolutions. The problem becomes harder with increasing values of the autoregression coefficient $\gamma$. Also, the performance of the spectral-NN estimator is comparable or better than the empirical estimator. The improvements are especially visible when the underlying functional observations are not smooth (Brownian sheet and Mat\'ern with lower values of $\nu$). The differences are more prominent in the 3D examples.

\begin{table}[t!]
\centering
\footnotesize
\caption{Relative error rates (in \%) of the empirical spectral density estimator (Emp) and the spectral-NN estimator (NN) in different 1D examples. The numbers are averages based on $25$ simulation runs. The corresponding standard errors are in the next line in italics and in a smaller font.\label{tab:1D_N_K}}
\bigskip

\setlength{\tabcolsep}{0.035in}
\begin{tabular}{r c rr c rr c rr c rr c rr c rr}
\multicolumn{19}{c}{Fixed AR coefficient $\gamma=0.5$, fixed resolution $K=200$, varying sample size $N$} \\ \hline
&& \multicolumn{2}{c}{} && \multicolumn{2}{c}{Integrated} && \multicolumn{11}{c}{Matern} \\
&& \multicolumn{2}{c}{Brownian Motion} && \multicolumn{2}{c}{Brownian Motion} && \multicolumn{2}{c}{$\nu=0.001$} && \multicolumn{2}{c}{$\nu=0.01$} && \multicolumn{2}{c}{$\nu=0.1$} && \multicolumn{2}{c}{$\nu=1$} \\
\cline{3-4} \cline{6-7} \cline{9-10} \cline{12-13} \cline{15-16} \cline{18-19}
$N$ && Emp & NN && Emp & NN && Emp & NN && Emp & NN && Emp & NN && Emp & NN \\ [4pt]
$100$	&& $38.10$	& $36.98$	&& $30.09$	& $30.60$	&& $254.28$	& $43.38$	&& $241.38$	& $50.45$	&& $82.34$	& $51.13$	&& $36.30$	& $36.45$\\
	&& $\se{1.58}$	& $\se{1.63}$	&& $\se{1.81}$	& $\se{1.96}$	&& $\se{1.03}$	& $\se{1.70}$	&& $\se{1.01}$	& $\se{1.60}$	&& $\se{1.11}$	& $\se{1.50}$	&& $\se{1.68}$	& $\se{1.69}$\\ [2pt]
$200$	&& $26.77$	& $25.72$	&& $20.77$	& $21.32$	&& $184.76$	& $33.88$	&& $175.20$	& $39.27$	&& $59.26$	& $38.01$	&& $25.30$	& $25.17$\\
	&& $\se{1.13}$	& $\se{1.03}$	&& $\se{1.28}$	& $\se{1.35}$	&& $\se{0.80}$	& $\se{1.08}$	&& $\se{0.79}$	& $\se{1.00}$	&& $\se{0.76}$	& $\se{0.96}$	&& $\se{1.18}$	& $\se{1.14}$\\ [2pt]
$400$	&& $19.51$	& $19.06$	&& $15.32$	& $15.42$	&& $134.16$	& $28.18$	&& $126.93$	& $30.29$	&& $42.35$	& $27.93$	&& $18.59$	& $18.58$\\
	&& $\se{0.90}$	& $\se{0.91}$	&& $\se{1.02}$	& $\se{1.09}$	&& $\se{0.42}$	& $\se{0.65}$	&& $\se{0.41}$	& $\se{0.72}$	&& $\se{0.51}$	& $\se{0.72}$	&& $\se{0.93}$	& $\se{0.92}$\\ [2pt]
$800$	&& $13.51$	& $13.30$	&& $10.40$	& $10.43$	&& $99.81$	& $24.47$	&& $94.03$	& $24.43$	&& $30.14$	& $20.98$	&& $12.90$	& $12.96$\\
	&& $\se{0.53}$	& $\se{0.54}$	&& $\se{0.64}$	& $\se{0.64}$	&& $\se{0.19}$	& $\se{0.51}$	&& $\se{0.18}$	& $\se{0.39}$	&& $\se{0.28}$	& $\se{0.38}$	&& $\se{0.54}$	& $\se{0.56}$\\ [2pt]
$1600$	&& $10.43$	& $10.37$	&& $8.30$	& $8.60$	&& $76.44$	& $23.61$	&& $71.52$	& $21.19$	&& $21.99$	& $17.05$	&& $10.03$	& $10.24$\\
	&& $\se{0.52}$	& $\se{0.53}$	&& $\se{0.63}$	& $\se{0.64}$	&& $\se{0.12}$	& $\se{0.49}$	&& $\se{0.11}$	& $\se{0.37}$	&& $\se{0.26}$	& $\se{0.32}$	&& $\se{0.54}$	& $\se{0.56}$ \\ [10pt]

\multicolumn{19}{c}{Fixed AR coefficient $\gamma=0.5$, fixed sample size $N=250$, varying resolution $K$} \\ \hline
&& \multicolumn{2}{c}{} && \multicolumn{2}{c}{Integrated} && \multicolumn{11}{c}{Matern} \\
&& \multicolumn{2}{c}{Brownian Motion} && \multicolumn{2}{c}{Brownian Motion} && \multicolumn{2}{c}{$\nu=0.001$} && \multicolumn{2}{c}{$\nu=0.01$} && \multicolumn{2}{c}{$\nu=0.1$} && \multicolumn{2}{c}{$\nu=1$} \\
\cline{3-4} \cline{6-7} \cline{9-10} \cline{12-13} \cline{15-16} \cline{18-19}
$K$ && Emp & NN && Emp & NN && Emp & NN && Emp & NN && Emp & NN && Emp & NN \\ [4pt]
$20$	&& $25.48$	& $24.90$	&& $21.01$	& $20.41$	&& $212.72$	& $72.74$	&& $199.33$	& $106.54$	&& $56.63$	& $42.03$	&& $24.18$	& $24.29$\\
	&& $\se{0.77}$	& $\se{0.75}$	&& $\se{0.83}$	& $\se{0.89}$	&& $\se{1.48}$	& $\se{3.00}$	&& $\se{1.40}$	& $\se{2.92}$	&& $\se{0.67}$	& $\se{0.79}$	&& $\se{0.81}$	& $\se{0.86}$\\ [2pt]
$40$	&& $27.03$	& $26.55$	&& $22.33$	& $22.34$	&& $188.81$	& $47.54$	&& $177.69$	& $63.38$	&& $55.65$	& $38.90$	&& $25.93$	& $25.86$\\
	&& $\se{1.30}$	& $\se{1.33}$	&& $\se{1.47}$	& $\se{1.51}$	&& $\se{1.00}$	& $\se{2.13}$	&& $\se{0.95}$	& $\se{1.71}$	&& $\se{0.85}$	& $\se{1.09}$	&& $\se{1.35}$	& $\se{1.37}$\\ [2pt]
$80$	&& $24.03$	& $23.24$	&& $18.66$	& $18.67$	&& $173.65$	& $34.39$	&& $163.93$	& $42.15$	&& $53.05$	& $34.86$	&& $22.82$	& $23.07$\\
	&& $\se{0.91}$	& $\se{0.97}$	&& $\se{1.03}$	& $\se{1.03}$	&& $\se{0.70}$	& $\se{1.03}$	&& $\se{0.68}$	& $\se{1.16}$	&& $\se{0.56}$	& $\se{0.82}$	&& $\se{0.92}$	& $\se{0.92}$\\ [2pt]
$160$	&& $24.20$	& $23.76$	&& $19.17$	& $19.34$	&& $166.56$	& $31.17$	&& $157.66$	& $35.90$	&& $52.39$	& $33.27$	&& $22.99$	& $23.00$\\
	&& $\se{0.93}$	& $\se{0.91}$	&& $\se{1.10}$	& $\se{1.06}$	&& $\se{0.60}$	& $\se{0.84}$	&& $\se{0.58}$	& $\se{0.72}$	&& $\se{0.51}$	& $\se{0.74}$	&& $\se{0.97}$	& $\se{0.98}$\\ [2pt]
$320$	&& $24.91$	& $24.24$	&& $19.64$	& $19.50$	&& $164.26$	& $30.85$	&& $155.89$	& $34.59$	&& $53.30$	& $34.40$	&& $23.84$	& $23.95$\\
	&& $\se{1.01}$	& $\se{1.02}$	&& $\se{1.17}$	& $\se{1.17}$	&& $\se{0.55}$	& $\se{0.91}$	&& $\se{0.54}$	& $\se{1.00}$	&& $\se{0.67}$	& $\se{0.98}$	&& $\se{1.05}$	& $\se{1.07}$\\ [2pt]
$640$	&& $24.83$	& $24.29$	&& $19.46$	& $19.71$	&& $163.55$	& $30.61$	&& $155.30$	& $34.11$	&& $53.58$	& $33.58$	&& $23.54$	& $23.46$\\
	&& $\se{1.49}$	& $\se{1.51}$	&& $\se{1.68}$	& $\se{1.66}$	&& $\se{0.74}$	& $\se{1.34}$	&& $\se{0.74}$	& $\se{1.32}$	&& $\se{1.01}$	& $\se{1.21}$	&& $\se{1.55}$	& $\se{1.58}$\\ [2pt]
$1280$	&& $25.19$	& $24.50$	&& $19.87$	& $19.99$	&& $162.13$	& $30.81$	&& $154.07$	& $33.56$	&& $53.20$	& $34.28$	&& $24.20$	& $24.37$\\
	&& $\se{0.94}$	& $\se{0.99}$	&& $\se{1.17}$	& $\se{1.15}$	&& $\se{0.48}$	& $\se{0.80}$	&& $\se{0.48}$	& $\se{0.76}$	&& $\se{0.58}$	& $\se{0.97}$	&& $\se{0.99}$	& $\se{1.03}$\\ [2pt]
$2560$	&& $23.86$	& $22.95$	&& $18.64$	& $18.36$	&& $161.14$	& $29.70$	&& $153.08$	& $32.18$	&& $52.20$	& $32.86$	&& $22.78$	& $22.69$\\
	&& $\se{0.99}$	& $\se{0.92}$	&& $\se{1.16}$	& $\se{1.17}$	&& $\se{0.42}$	& $\se{0.83}$	&& $\se{0.43}$	& $\se{0.79}$	&& $\se{0.59}$	& $\se{0.85}$	&& $\se{1.04}$	& $\se{1.02}$
\end{tabular}
\end{table}

\begin{table}[t!]
\centering
\footnotesize
\caption{Relative error rates (in \%) of the empirical spectral density estimator (Emp) and the spectral-NN estimator (NN) in different 1D examples with a fixed sample size $N=250$, fixed resolution $K=200$ and varying AR coefficient $\gamma$. The numbers are averages based on $25$ simulation runs. The corresponding standard errors are in the next line in italics and in a smaller font.\label{tab:1D_gamma}}

\setlength{\tabcolsep}{0.035in}
\begin{tabular}{r c rr c rr c rr c rr c rr c rr}
&& \multicolumn{2}{c}{} && \multicolumn{2}{c}{Integrated} && \multicolumn{11}{c}{Matern} \\
&& \multicolumn{2}{c}{Brownian Motion} && \multicolumn{2}{c}{Brownian Motion} && \multicolumn{2}{c}{$\nu=0.001$} && \multicolumn{2}{c}{$\nu=0.01$} && \multicolumn{2}{c}{$\nu=0.1$} && \multicolumn{2}{c}{$\nu=1$} \\
\cline{3-4} \cline{6-7} \cline{9-10} \cline{12-13} \cline{15-16} \cline{18-19}
$\gamma$ && Emp & NN && Emp & NN && Emp & NN && Emp & NN && Emp & NN && Emp & NN \\ [4pt]
$0.1$	&& $23.84$	& $23.22$	&& $18.31$	& $18.39$	&& $166.03$	& $30.26$	&& $157.32$	& $34.45$	&& $52.80$	& $33.11$	&& $22.77$	& $22.76$\\
	&& $\se{0.82}$	& $\se{0.85}$	&& $\se{1.02}$	& $\se{1.00}$	&& $\se{0.51}$	& $\se{0.71}$	&& $\se{0.51}$	& $\se{0.68}$	&& $\se{0.50}$	& $\se{0.68}$	&& $\se{0.84}$	& $\se{0.85}$\\ [2pt]
$0.25$	&& $24.39$	& $23.68$	&& $18.88$	& $19.13$	&& $166.23$	& $31.05$	&& $157.53$	& $35.18$	&& $53.20$	& $33.64$	&& $23.34$	& $23.43$\\
	&& $\se{1.00}$	& $\se{0.97}$	&& $\se{1.21}$	& $\se{1.22}$	&& $\se{0.59}$	& $\se{0.91}$	&& $\se{0.59}$	& $\se{0.84}$	&& $\se{0.62}$	& $\se{0.76}$	&& $\se{1.02}$	& $\se{1.04}$\\ [2pt]
$0.5$	&& $25.80$	& $25.12$	&& $20.36$	& $20.66$	&& $166.82$	& $33.23$	&& $158.14$	& $37.73$	&& $54.21$	& $34.69$	&& $24.76$	& $24.72$\\
	&& $\se{1.40}$	& $\se{1.42}$	&& $\se{1.64}$	& $\se{1.64}$	&& $\se{0.81}$	& $\se{1.23}$	&& $\se{0.80}$	& $\se{1.18}$	&& $\se{0.88}$	& $\se{1.14}$	&& $\se{1.43}$	& $\se{1.53}$\\ [2pt]
$0.75$	&& $30.24$	& $29.87$	&& $25.36$	& $25.71$	&& $170.14$	& $37.77$	&& $161.42$	& $43.20$	&& $57.47$	& $39.25$	&& $29.23$	& $29.28$\\
	&& $\se{1.87}$	& $\se{1.96}$	&& $\se{2.10}$	& $\se{2.16}$	&& $\se{1.46}$	& $\se{1.65}$	&& $\se{1.40}$	& $\se{1.52}$	&& $\se{1.25}$	& $\se{1.54}$	&& $\se{1.91}$	& $\se{1.96}$\\ [2pt]
$0.9$	&& $49.60$	& $48.91$	&& $42.33$	& $42.54$	&& $193.18$	& $57.66$	&& $183.83$	& $61.94$	&& $74.75$	& $57.74$	&& $48.72$	& $48.72$\\
	&& $\se{1.60}$	& $\se{1.64}$	&& $\se{1.48}$	& $\se{1.44}$	&& $\se{4.74}$	& $\se{1.63}$	&& $\se{4.48}$	& $\se{1.62}$	&& $\se{1.47}$	& $\se{1.40}$	&& $\se{1.62}$	& $\se{1.66}$
\end{tabular}
\end{table}
\begin{table}[h!]
\centering
\footnotesize
\caption{Average computing times (in seconds) and maximum memory usage (in MB) of the empirical spectral density estimator (Emp) and the spectral-NN estimator (NN) in different 1D examples. For NN, computing times with GPU are shown in the next line in italics. The codes were run on a computer with 64 GiB RAM, AMD Ryzen 9 5900X (3.7 GHz) CPU, NVIDIA GeForce RTX 3090 GPU, and Ubuntu 24.04.2 LTS (64-bit) OS.\label{tab:1D_time_memory}}

\setlength{\tabcolsep}{0.05in}
\begin{tabular}{l c rr c rr c rr c rr c rr}
\multicolumn{16}{c}{Fixed resolution $K=200$, variying sample size $N$.} \\\hline
$N$ && \multicolumn{2}{c}{$100$} && \multicolumn{2}{c}{$200$} && \multicolumn{2}{c}{$400$} && \multicolumn{2}{c}{$800$} && \multicolumn{2}{c}{$1600$} \\
 &&	Emp & NN && Emp & NN && Emp & NN && Emp & NN && Emp & NN \\ [3pt]
Fit && $0.12$ & $113.92$ && $0.26$ & $164.85$ && $0.53$ & $217.73$ && $1.03$ & $341.88$ && $2.07$ & $2798.25$ \\
    && & $\gpu{245.14}$ && & $\gpu{269.01}$ && & $\gpu{315.34}$ && & $\gpu{406.07}$ && & $\gpu{623.84}$ \\
Eval  && $37.81$ & $101.50$ && $36.20$ & $100.57$ && $36.04$ & $100.90$ && $35.84$ & $101.38$ && $37.40$ & $100.35$ \\
      && & $\gpu{1.72}$ && & $\gpu{1.72}$ && & $\gpu{1.72}$ && & $\gpu{1.72}$ && & $\gpu{1.75}$ \\
Total && $37.93$ & $215.42$ && $36.46$ & $265.42$ && $36.57$ & $318.63$ && $36.87$ & $443.26$ && $39.47$ & $2898.61$ \\
      && & $\gpu{246.86}$ && & $\gpu{270.73}$ && & $\gpu{317.06}$ && & $\gpu{407.79}$ && & $\gpu{625.59}$ \\ [2pt]
Memory && $121$ & $689$ && $114$ & $694$ && $121$ & $695$ && $120$ & $706$ && $121$ & $699$ \\ [3pt]

\multicolumn{16}{c}{Fixed sample size $N=250$, variying resolution $K$.} \\\hline
$K$ && \multicolumn{2}{c}{$160$} && \multicolumn{2}{c}{$320$} && \multicolumn{2}{c}{$640$} && \multicolumn{2}{c}{$1280$} && \multicolumn{2}{c}{$2560$} \\
 &&	Emp & NN && Emp & NN && Emp & NN && Emp & NN && Emp & NN \\ [3pt]
Fit && $0.26$ & $172.19$ && $0.34$ & $188.46$ && $0.40$ & $221.23$ && $0.72$ & $265.94$ && $7.96$ & $403.10$ \\
    && & $\gpu{279.26}$ && & $\gpu{277.88}$ && & $\gpu{277.73}$ && & $\gpu{277.94}$ && & $\gpu{280.23}$ \\
Eval  && $34.28$ & $97.75$ && $40.08$ & $97.80$ && $38.08$ & $99.63$ && $41.39$ & $99.05$ && $55.02$ & $98.31$ \\
      && & $\gpu{1.72}$ && & $\gpu{1.73}$ && & $\gpu{1.73}$ && & $\gpu{1.73}$ && & $\gpu{1.73}$ \\
Total && $34.54$ & $269.95$ && $40.42$ & $286.26$ && $38.48$ & $320.86$ && $42.11$ & $364.99$ && $62.98$ & $501.41$ \\
      && & $\gpu{280.98}$ && & $\gpu{279.61}$ && & $\gpu{279.46}$ && & $\gpu{279.67}$ && & $\gpu{281.96}$ \\ [2pt]
Memory && $120$ & $695$ && $140$ & $688$ && $190$ & $689$ && $396$ & $691$ && $1235$ & $698$
\end{tabular}
\end{table}

\begin{table}[t!]
\centering
\footnotesize
\caption{Relative error rates (in \%) of the empirical spectral density estimator (Emp) and the spectral-NN estimator (NN) in different 3D examples. The numbers are averages based on $25$ simulation runs. The corresponding standard errors are in the next line in italics and in a smaller font. A dash (---) indicates that the program failed due to insufficient memeory.\label{tab:3D_N_K}}
\bigskip

\setlength{\tabcolsep}{0.03in}
\begin{tabular}{r c rr c rr c rr c rr c rr c rr}
\multicolumn{19}{c}{Fixed AR coefficient $\gamma=0.5$, fixed resolution $K=15$, varying sample size $N$} \\ \hline
&& \multicolumn{2}{c}{} && \multicolumn{2}{c}{Integrated} && \multicolumn{11}{c}{Matern} \\
&& \multicolumn{2}{c}{Brownian Sheet} && \multicolumn{2}{c}{Brownian Sheet} && \multicolumn{2}{c}{$\nu=0.001$} && \multicolumn{2}{c}{$\nu=0.01$} && \multicolumn{2}{c}{$\nu=0.1$} && \multicolumn{2}{c}{$\nu=1$} \\
\cline{3-4} \cline{6-7} \cline{9-10} \cline{12-13} \cline{15-16} \cline{18-19}
$N$ && Emp & NN && Emp & NN && Emp & NN && Emp & NN && Emp & NN && Emp & NN \\ [4pt]
$100$	&& $56.05$	& $46.77$	&& $33.95$	& $36.86$	&& $15348.49$	& $202.79$	&& $13121.01$	& $162.75$	&& $535.52$	& $98.74$	&& $50.49$	& $48.80$\\
	&& $\se{0.87}$	& $\se{1.07}$	&& $\se{1.59}$	& $\se{1.65}$	&& $\se{33.24}$	& $\se{10.13}$	&& $\se{21.91}$	& $\se{5.45}$	&& $\se{1.46}$	& $\se{2.75}$	&& $\se{0.83}$	& $\se{1.11}$\\ [2pt]
$200$	&& $43.19$	& $37.20$	&& $28.12$	& $27.31$	&& $11046.12$	& $216.70$	&& $9495.02$	& $157.97$	&& $387.94$	& $81.10$	&& $38.57$	& $37.27$\\
	&& $\se{0.78}$	& $\se{0.89}$	&& $\se{1.34}$	& $\se{1.26}$	&& $\se{15.06}$	& $\se{7.11}$	&& $\se{8.57}$	& $\se{5.62}$	&& $\se{1.00}$	& $\se{1.97}$	&& $\se{1.25}$	& $\se{1.30}$\\ [2pt]
$400$	&& $31.57$	& $29.24$	&& $22.35$	& $21.27$	&& $7902.86$	& $361.22$	&& $6794.62$	& $177.41$	&& $280.35$	& $63.53$	&& $28.67$	& $28.01$\\
	&& $\se{0.65}$	& $\se{0.92}$	&& $\se{0.54}$	& $\se{0.86}$	&& $\se{8.48}$	& $\se{9.30}$	&& $\se{5.22}$	& $\se{4.23}$	&& $\se{0.81}$	& $\se{1.70}$	&& $\se{0.54}$	& $\se{0.56}$\\ [2pt]
$800$	&& $22.48$	& $20.64$	&& $18.33$	& $18.37$	&& $5676.39$	& $307.82$	&& $4874.80$	& $168.72$	&& $203.10$	& $52.58$	&& $19.98$	& $20.10$\\
	&& $\se{0.26}$	& $\se{0.37}$	&& $\se{0.34}$	& $\se{1.59}$	&& $\se{4.59}$	& $\se{9.53}$	&& $\se{3.04}$	& $\se{2.46}$	&& $\se{0.36}$	& $\se{0.84}$	&& $\se{0.41}$	& $\se{0.50}$\\ [2pt]
$1600$	&& $17.93$	& $17.18$	&& $17.66$	& $17.17$	&& $4090.39$	& $265.60$	&& $3519.45$	& $163.90$	&& $150.16$	& $46.29$	&& $13.95$	& $14.41$\\
	&& $\se{0.28}$	& $\se{0.37}$	&& $\se{0.32}$	& $\se{1.66}$	&& $\se{1.91}$	& $\se{6.19}$	&& $\se{2.33}$	& $\se{3.37}$	&& $\se{0.27}$	& $\se{0.44}$	&& $\se{0.28}$	& $\se{0.27}$\\ [10pt]

\multicolumn{19}{c}{Fixed AR coefficient $\gamma=0.5$, fixed sample size $N=250$, varying resolution $K$} \\ \hline
&& \multicolumn{2}{c}{} && \multicolumn{2}{c}{Integrated} && \multicolumn{11}{c}{Matern} \\
&& \multicolumn{2}{c}{Brownian Sheet} && \multicolumn{2}{c}{Brownian Sheet} && \multicolumn{2}{c}{$\nu=0.001$} && \multicolumn{2}{c}{$\nu=0.01$} && \multicolumn{2}{c}{$\nu=0.1$} && \multicolumn{2}{c}{$\nu=1$} \\
\cline{3-4} \cline{6-7} \cline{9-10} \cline{12-13} \cline{15-16} \cline{18-19}
$K$ && Emp & NN && Emp & NN && Emp & NN && Emp & NN && Emp & NN && Emp & NN \\ [4pt]
$10$	&& $40.14$	& $35.35$	&& $31.96$	& $41.57$	&& $10015.70$	& $362.65$	&& $8631.20$	& $287.94$	&& $354.54$	& $96.48$	&& $34.84$	& $33.37$\\
	&& $\se{0.71}$	& $\se{0.96}$	&& $\se{0.93}$	& $\se{3.89}$	&& $\se{11.61}$	& $\se{10.03}$	&& $\se{12.17}$	& $\se{6.60}$	&& $\se{1.07}$	& $\se{2.13}$	&& $\se{0.63}$	& $\se{0.68}$\\ [2pt]
$15$	&& $37.66$	& $32.18$	&& $24.62$	& $30.96$	&& $9937.27$	& $234.32$	&& $8517.69$	& $176.77$	&& $348.75$	& $71.70$	&& $34.39$	& $32.66$\\
	&& $\se{0.62}$	& $\se{0.73}$	&& $\se{0.78}$	& $\se{1.76}$	&& $\se{10.27}$	& $\se{6.99}$	&& $\se{12.25}$	& $\se{5.20}$	&& $\se{0.96}$	& $\se{1.88}$	&& $\se{0.74}$	& $\se{0.73}$\\ [2pt]
$20$	&& $38.37$	& $32.83$	&& $24.11$	& $37.88$	&& $9899.69$	& $168.16$	&& $8492.14$	& $118.38$	&& $346.48$	& $65.13$	&& $33.92$	& $32.45$\\
	&& $\se{0.85}$	& $\se{0.88}$	&& $\se{1.51}$	& $\se{3.32}$	&& $\se{6.84}$	& $\se{7.29}$	&& $\se{8.58}$	& $\se{2.30}$	&& $\se{0.73}$	& $\se{1.35}$	&& $\se{0.80}$	& $\se{0.80}$\\ [2pt]
$25$	&& $38.21$	& $33.09$	&& $21.49$	& $33.19$	&& $9878.38$	& $153.95$	&& $8503.70$	& $95.31$	&& $345.88$	& $60.11$	&& $33.50$	& $32.02$\\
	&& $\se{0.51}$	& $\se{0.65}$	&& $\se{0.91}$	& $\se{3.09}$	&& $\se{9.79}$	& $\se{10.05}$	&& $\se{8.41}$	& $\se{3.10}$	&& $\se{0.71}$	& $\se{0.97}$	&& $\se{0.55}$	& $\se{0.60}$\\ [2pt]
$30$	&& ---	& $33.53$	&& ---	& $33.94$	&& ---	& $143.89$	&& ---	& $80.50$	&& ---	& $57.34$	&& ---	& $32.18$\\
	&& 	& $\se{0.73}$	&& 	& $\se{2.75}$	&& 	& $\se{7.92}$	&& 	& $\se{2.52}$	&& 	& $\se{0.85}$	&& 	& $\se{0.65}$
\end{tabular}
\end{table}

\begin{table}[!t]
\centering
\footnotesize
\caption{Average computing times (in seconds) and maximum memory usage (in MB) of the empirical spectral density estimator (Emp) and the spectral-NN estimator (NN) in different 3D examples. For NN, computing times with GPU are shown in the next line in italics. The codes were run on a computer with 64 GiB RAM, AMD Ryzen 9 5900X (3.7 GHz) CPU, NVIDIA GeForce RTX 3090 GPU, and Ubuntu 24.04.2 LTS (64-bit) OS. A dash (---) indicates that the program failed due to insufficient memeory.\label{tab:3D_time_memory}}
\bigskip

\setlength{\tabcolsep}{0.04in}
\begin{tabular}{l c rr c rr c rr c rr c rr}
\multicolumn{16}{c}{Fixed resolution $K=15$, variying sample size $N$.} \\\hline
$N$ && \multicolumn{2}{c}{$100$} && \multicolumn{2}{c}{$200$} && \multicolumn{2}{c}{$400$} && \multicolumn{2}{c}{$800$} && \multicolumn{2}{c}{$1600$} \\
 &&	Emp & NN && Emp & NN && Emp & NN && Emp & NN && Emp & NN \\ [3pt]
Fit && $33.72$ & $338.60$ && $55.45$ & $432.08$ && $141.82$ & $568.61$ && $290.24$ & $902.78$ && $596.87$ & $2362.70$ \\
    && & $\gpu{247.69}$ && & $\gpu{271.48}$ && & $\gpu{316.58}$ && & $\gpu{405.20}$ && & $\gpu{627.75}$ \\
Eval  && $578.62$ & $100.09$ && $578.23$ & $98.39$ && $573.87$ & $99.27$ && $579.75$ & $101.93$ && $580.84$ & $103.20$ \\
      && & $\gpu{1.97}$ && & $\gpu{1.97}$ && & $\gpu{1.97}$ && & $\gpu{1.96}$ && & $\gpu{2.00}$ \\
Total && $612.34$ & $438.69$ && $633.68$ & $530.46$ && $715.69$ & $667.88$ && $869.99$ & $1004.71$ && $1177.71$ & $2465.90$ \\
      && & $\gpu{249.66}$ && & $\gpu{273.45}$ && & $\gpu{318.55}$ && & $\gpu{407.16}$ && & $\gpu{629.75}$ \\ [2pt]
Memory && $2068$ & $674$ && $2069$ & $699$ && $2071$ & $696$ && $2074$ & $698$ && $2080$ & $754$ \\ [10pt]

\multicolumn{16}{c}{Fixed sample size $N=250$, variying resolution $K$.} \\\hline
$K$ && \multicolumn{2}{c}{$10$} && \multicolumn{2}{c}{$15$} && \multicolumn{2}{c}{$20$} && \multicolumn{2}{c}{$25$} && \multicolumn{2}{c}{$30$} \\
 &&	Emp & NN && Emp & NN && Emp & NN && Emp & NN && Emp & NN \\ [3pt]
Fit && $0.50$ & $224.34$ && $85.94$ & $465.16$ && $578.62$ & $873.62$ && $2292.85$ & $1522.48$ && --- & $2782.41$ \\
    && & $\gpu{280.72}$ && & $\gpu{279.68}$ && & $\gpu{286.78}$ && & $\gpu{292.13}$ && & $\gpu{317.63}$ \\
Eval  && $190.34$ & $97.01$ && $578.56$ & $96.68$ && $1402.81$ & $99.14$ && $2236.63$ & $98.02$ && --- & $102.75$ \\
      && & $\gpu{1.98}$ && & $\gpu{1.96}$ && & $\gpu{1.97}$ && & $\gpu{1.97}$ && & $\gpu{1.97}$ \\
Total && $190.84$ & $321.35$ && $664.50$ & $561.84$ && $1981.44$ & $972.76$ && $4529.48$ & $1620.50$ && --- & $2885.16$ \\
      && & $\gpu{282.70}$ && & $\gpu{281.64}$ && & $\gpu{288.75}$ && & $\gpu{294.10}$ && & $\gpu{319.60}$ \\ [2pt]
Memory && $294$ & $699$ && $2070$ & $698$ && $11090$ & $850$ && $41961$ & $1433$ && --- & $3162$
\end{tabular}
\end{table}

In terms of the computing times and memory requirements, we again see the usefulness of the spectral-NN estimator, particularly when the resolution of the data is high, especially in the 3D examples. Even at a moderate resolution of $15 \times 15 \times 15$, the empirical estimator requires almost thrice the memory required by the spectral-NN estimator. The computing time for the empirical estimator also increases exponentially, with almost thrice that of the spectral-NN estimator at a resolution of $25 \times 25 \times 25$. Moreover, we observe substantial reduction in computing time for spectral-NN with GPU computing, especially at large resolutions. At a resolution of $20 \times 20 \times 20$, spectral-NN with GPU requires less than one-sixth of the time required by the empirical estimator. This becomes even more substantial at a resolution of $25 \times 25 \times 25$, where spectral-NN with GPU requires less than one-fifteenth of the time required by the empirical estimator. At this resolution, the empirical estimator requires almost $42$ gigabytes of memory, which is much more than what is found on a regular computer. This is in stark contrast to the less than $1.5$ gigabytes of memory required by the spectral-NN estimator. At a resolution of $30 \times 30 \times 30$, the empirical estimator completely breaks down, requiring more than $100$ gigabytes of memory. In comparison, the spectral-NN estimator requires only $3.2$ gigabytes of memory, which is easily available on most regular computers.

\end{document}